\documentclass[twocolumn,floatfix, aps, prd, a4paper, 10pt, showpacs, superscriptaddress, nofootinbib, reprint]{revtex4-2}
\tighten
\usepackage{amsmath}
\usepackage{color}
\usepackage{xcolor}
\usepackage{graphicx}
\usepackage{caption}
\usepackage{acronym}
\usepackage[justification=justified]{caption}
\usepackage{subcaption}
\usepackage{dcolumn}
\usepackage{bm}
\usepackage{stackengine}

\usepackage{enumitem,amssymb}
\usepackage{hyperref}
\usepackage[normalem]{ulem}

\makeatletter
\count@=`A \advance\count@\m@ne
\@whilenum\count@<`Z\do{%
  \advance\count@\@ne
  \begingroup\uccode`a=\count@
  \uppercase{\endgroup\DeclareMathSymbol{a}}{\mathalpha}{operators}{\count@}%
}
\makeatother

\makeatletter
\count@=`a \advance\count@\m@ne
\@whilenum\count@<`z\do{%
  \advance\count@\@ne
  \begingroup\uccode`a=\count@
  \uppercase{\endgroup\DeclareMathSymbol{a}}{\mathalpha}{operators}{\count@}%
}
\makeatother

\usepackage{ragged2e} 

\makeatletter
\renewcommand{\@makecaption}[2]{%
  \vskip\abovecaptionskip
  \sbox\@tempboxa{\RaggedRight\small\textbf{#1:} #2}%
  \ifdim \wd\@tempboxa > \hsize
    \RaggedRight\small\textbf{#1:} #2\par
  \else
    \hb@xt@\hsize{\hfil\box\@tempboxa\hfil}%
  \fi
  \vskip\belowcaptionskip
}
\makeatother

\newcommand{\ie}{\textit{i.e. }}

\newcommand{\As}{A$\# ~$}
\definecolor{darkpink}{RGB}{231, 84, 128}

\definecolor{cadmiumgreen}{rgb}{0.0, 0.42, 0.24}

\def\mycmd{1}

\begin{document}

\preprint{eos-model-selection}
\title{Optimizing Bayesian model selection for equation of state of cold neutron stars}

\author{Rahul Kashyap}
\email{rahulkashyap@iitb.ac.in}
\affiliation{Department of Physics, Indian Institute of Technology Bombay, Mumbai 400076, India}
\affiliation{Institute for Gravitation and the Cosmos and Physics Department, Penn State University, University Park PA 16802, USA}
\affiliation{Department of Astronomy and Astrophysics, Penn State University, University Park PA 16802, USA}
\author{Ish Gupta}
\affiliation{Institute for Gravitation and the Cosmos and Physics Department, Penn State University, University Park PA 16802, USA}
\affiliation{Department of Astronomy and Astrophysics, Penn State University, University Park PA 16802, USA}
\author{Arnab Dhani}
\affiliation{Max-Planck Institute for Gravitational Physics, Albert-Einstein Institute, Am M\"uhlenberg 1, 14476 Potsdam, Germany}
\author{Monica Bapna}
\address{International Centre for Theoretical Sciences, Tata Institute of Fundamental Research, Bengaluru 560089, India}
\author{Bangalore Sathyaprakash}
\affiliation{Institute for Gravitation and the Cosmos and Physics Department, Penn State University, University Park PA 16802, USA}
\affiliation{Department of Astronomy and Astrophysics, Penn State University, University Park PA 16802, USA}
\affiliation{School of Physics and Astronomy, Cardiff University, Cardiff, CF24 3AA, United Kingdom}

\date{\today}

\begin{abstract}
We introduce a computational framework, \ac{BEOMS} to evaluate multiple Bayesian model selection methods in the context of determining the \ac{EOS} for cold \ac{NS}, focusing on their performance with current and next-generation \ac{GW} observatories. We conduct a systematic comparison of various \ac{EOS} models by using posterior distributions obtained from EOS-agnostic Bayesian inference of binary parameters applied to \acp{GW} from a population of \ac{BNS} mergers. The cumulative evidence for each model is calculated in a multi-dimensional parameter space characterized by neutron star masses and tidal deformabilities. Our findings indicate that Bayesian model selection is most effective when performed in the two-dimensional subspace of component mass and tidal deformability, requiring fewer events to distinguish between EOS models with high confidence. Furthermore, we establish a relationship between the precision of tidal deformability measurements and the accuracy of model selection, taking into account the evolving sensitivities of current and planned GW observatories. \ac{BEOMS} offers computational efficiency and can be adapted to execute model selection for gravitational wave data from other sources.
\end{abstract}

\maketitle

\section{\label{sec:intro}Introduction}

\ac{GW} astronomy has revolutionized our ability to probe the ultradense matter found in \ac{NS} cores \citep{LIGOScientific:2014pky, VIRGO:2014yos, LIGOScientific:2018hze, LIGOScientific:2017vwq, LIGOScientific:2018mvr}. One of the key observables linking \ac{GW} signals to the internal composition of \acp{NS} is the tidal deformability parameter ($\Lambda$), that quantifies the extent to which an NS’s shape deforms under the influence of its companion’s tidal field  \citep{Lai:1993pa, Cutler:1994ys, Kokkotas:1999bd, Flanagan:2007ix, Hinderer:2009ca}. The tidal deformability is directly related to the underlying \ac{EOS} of the NS through a unique mapping between the pressure-density ($p$-$\epsilon$) curve of cold nuclear matter and the tidal deformability-mass ($\Lambda$-$m$) relation \cite{Lindblom1992-rg}, which is derived from solving the first-order general relativisitic perturbations of matter in NSs.

In the context of GW astronomy, the GW signals from \ac{BNS} mergers carry imprints of the component masses and tidal deformabilities. These parameters can be extracted from the waveform’s amplitude and phase evolution, especially during the late inspiral phase, where tidal effects become prominent. For precise characterization in the pre-merger regime, waveform models are typically calibrated to high-fidelity numerical relativity simulations of BNS mergers, capturing the non-linear interactions between the two stars \cite{Hinderer:2016eia, Dietrich:2017feu, Dietrich:2017aum, Dietrich:2018uni, Dietrich:2019kaq, Nagar:2018zoe, Henry:2020pzq, Henry:2020ski}. Bayesian inference techniques are used to estimate both intrinsic parameters (masses, spins, tidal deformabilities) and extrinsic parameters (sky position, inclination, distance) of the system from the GW data. However, the accuracy of these measurements is sensitive to the presence of noise artifacts and the choice of priors, which can introduce systematic biases in the posterior distributions \citep{Gamba:2020wgg, Vitale:2017cfs}.

To mitigate these uncertainties and improve the ``evidence" for a model, it is essential to combine data from multiple independent BNS events. This cumulative approach allows for a more robust measurement of shared astrophysical properties, such as the EOS of cold dense matter, and improves the confidence in population-wide characteristics. Such methods are also critical for testing alternative theories of gravity and constraining the distribution of BNS systems \cite{Mandel:2009pc, zhu2018inferring, Vitale2020-hj}. In this paper, we combine measurements from multiple simulated BNS GW events and compare various Bayesian model selection techniques to determine the optimal strategy for the inference of the EOS, proposing an approach that maximizes the effectiveness of model selection for the state of matter in NS cores.

There are four principal approaches in the literature for Bayesian model selection of the dense matter EOS. In the first approach, evidence for different EOSs is obtained by directly measuring appropriately chosen physical parameters, such as their \emph{polytropic} \citep{Read:2008iy} or \emph{spectral} \citep{Lindblom2010-bp} indices and the sound speed \citep{Abdelsalhin2018-yb, Alsing2018-sl, Ray2023-gx}. In the second approach, EOSs are compared against each other by computing the evidence for EOS models represented by $\Lambda$-$m$ curves \citep{Agathos:2015uaa, LIGOScientific:2019eut, Ghosh:2021eqv, Biswas:2021pvm}. For instance, \citet{Markakis2009-da} provide correlations using numerical relativity simulations of the inspiral and postmerger GW signals from BNS mergers that can be used to constrain the EOS. 

\citet{Agathos:2015uaa}, among the first to employ a Bayesian pipeline for EOS model selection, computed pairwise Bayes factors for each EOS and a collection of simulated BNS events. However, they approximated the tidal deformability of NSs by a Taylor expansion around a fiducial mass. 

\citet{Legred:2021hdx} showed that artificial correlations appear purely due to our parameterization of EOSs using some phenomenological, though physically motivated, parameters such as spectral indices, sound speed or piecewise polytropic indices. This raises concerns over the correct approach to find the basic degrees of freedom in the problem of dense-matter EOS. The first-principle quantum many-body theory of QCD matter is yet to establish the basic degrees of freedom at any specific density. However, such a theory must align with NS matter at low densities, such as nuclear to saturation density. This viewpoint has led to the notion in the literature that the parameters of this theory should be corroborated with GW observations from BNS mergers, thereby connecting first-principles calculations with GW observations \citep{Margueron:2017eqc, Iacovelli2023-yg}. In the third approach, \citet{Essick:2019ldf} and \citet{Landry:2018prl} circumvent difficulties in the parameterized methods by applying a non-parametric technique for the selection of the EOS models. Finally, the fourth approach relies on EOS-agnostic Bayesian inference to find the relative evidence (i.e., the Bayes factor) for a collection of EOS models \citep{LIGOScientific:2019eut, Ghosh:2021eqv}.

In this study, we adopt the model selection framework introduced in Ref.~\citep{LIGOScientific:2019eut} and extend its application to a population of simulated binary neutron star (BNS) signals expected from future gravitational-wave (GW) detectors with enhanced sensitivities. In addition, we choose to compute and compare the Bayesian evidence for different equations of state (EOSs) not only in the full four-dimensional parameter space defined by the component masses and their tidal deformabilities $(m_1$-$m_2$-$\Lambda_1$-$\Lambda_2)$, but also within several lower-dimensional subspaces, namely $(m_1-\Lambda_1)$, $(m_2-\Lambda_2)$, and $(\tilde{\Lambda}-\eta)$. By systematically evaluating the performance of EOS inference across these distinct subspaces, we are able to identify the most effective strategy for recovering the true EOS using the fourth model selection approach. We stress that this targeted choice of subspaces significantly improves the computational efficiency of the model selection procedure while preserving the essential information required to distinguish between different EOSs.

Given data which a model should explain, Bayesian evidence penalizes complex models and models with more parameters, technically known as \emph{Occam's penalty}. There are two known ways of enforcing Occam's penalty. In the first case, one favors a model that conforms to a subjective prior before performing Bayesian inference. The other approach is naturally included in Bayesian inference, where the computation of evidence in a higher-dimensional space, corresponding to models with greater number of parameters, leads to a lower evidence \citep{Mackay2003, Jefferys1992-qq}. We incorporate the subjective prior by using EOS-dependent maximum mass limits through a reweighting of the posterior distribution that is initially EOS-agnostic and is obtained under the assumption of an arbitrarily large prior maximum mass \cite{Legred:2022pyp}. Furthermore, for the evidence calculation we adopt an astrophysically motivated, reduced-dimensional parameter space, which helps ensure that any loss in evidence remains minimal when compared to carrying out the computation in the full parameter space.

\citet{Kastaun:2019bxo} have shown the impact of prior and model assumptions on detection of finite tidal effects, as well as EOS model selection. They have argued that credible intervals on tidal parameters cannot be used to rule out any EOS model, rather Bayesian model selection must be used. We follow this path in the present work. 

In addition, we do not employ any specific parameterization of EOS in our collection, which allows us to perform model selection on any given mass-tidal deformability curve ($m$-$\Lambda$) or, equivalently, pressure-density ($p$-$\epsilon$) curve. Since GW observations give us macroscopic information about the deformability of NSs, we dissociate the work of connecting to microscopic degrees of EOS ($p$-$\epsilon$) and present model selection results in the $m$-$\Lambda$ plane regarding which uncertainties are minimal.

Observations of the X-ray Pulsar using NICER lead to independent constraints on NS-EOS, which could be combined with GW observations to put tighter constraints on NS-EOS \cite{Raaijmakers:2019dks, Raaijmakers:2021uju, Raaijmakers2019-tg, Miller:2021qha, Bogdanov2022-nn, Bogdanov2019-le, Miller2019-ll}. \citealt{Ascenzi2024-bg} provide summary and future prospects of constraining EOS using radio, optical, X-ray and $\gamma-$ray observations of NSs. \citealt{Koehn2024-mz} offer a commendable recent overview of constraints imposed by multimessenger observations as well as nuclear physics from both experimental and theoretical perspectives. They integrate findings from heavy ion collisions and PREX experiments with the theoretical framework of $\chi$EFT and pQCD to address the EOS at lower and upper limits of neutron star densities.

Constraining deviations from quasi-universal relations (qUR) has been one of the popular methods to reduce the parameter space in constraining EOS \citep{LIGOScientific:2018cki, Legred:2023als}. \citet{Kastaun:2019bxo} argued against the use of universal relations to place a lower bound on tidal deformability. \citet{Kashyap:2022wzr} showed that in the $m_1$-$m_2$ space systematic biases from qUR could be significant and bias the estimate of the correct EOS. They addressed this issue by proposing correction for the systematic errors at the time of model selection. Their $\chi^2$ technique \citep{Kashyap:2022wzr} for EOS model selection is found to be ineffective when applied to BNS observations by present and near-future GW detectors. This is simply due to large statistical errors that result in the denominator of the expression for $\chi^2$ (see Eq. (12) of \citet{Kashyap:2022wzr}). To mitigate the aforementioned problems, we extend our work to the calculation of Bayesian evidence for a collection of EOS models in different spaces, thereby evaluating their comparative efficiencies. We validate our work by presenting the injection studies of BNS mergers in three different current and planned gravitational wave detector networks.

We defer the study of the impact of eccentricity to future analyses. \citealt{roy2024impact} has talked about the impact of eccentricity on the tidal deformability and concluded that it could have a large impact for BNS events observed by the XG detectors when the signal is analyzed from 10 Hz. We assume NSs to have finite, but small, spins which implies spin-induced quadrupole moments are sub-dominant effects in the waveform models. We have not looked at the impact of waveform uncertainties on our model selection results. We refer to Refs. \cite{Samajdar:2018dcx, Kastaun:2019bxo, Samajdar:2019ulq, Gamba:2020wgg} for such discussions. \citealt{Rreali2023impact} presents biases due to astrophysical confusion noise and found significant biases even for high signal-to-noise ratio (SNR) BNS mergers observed in next-generation detectors (however, also see \citep{Gupta:2024lft} which finds the effect of foreground noise on parameter estimation to be minimal). It has also been argued that NS oscillations with timescales of the order of binary period could lead to biased estimate of NS tidal deformability and hence the EOS \citep{Pradhan2023-ua, Pratten2021-re, Gupta2023-ec}. They mention the relative bias to be maximum of 50\% for intermediate mass NS which is within the statistical uncertainty for current and near future configurations of \ac{GW} detectors.

In the remainder of the paper, we outline the details of our calculations and results. In Sec.~\ref{sec:method}, we describe the calculation of evidence in various spaces involving transformations and marginalization. We present the distribution of evidence in different space for three EOS and three detector configurations. We then outline the comparison of cumulative evidence of EOS models in Sec~\ref{sec:Results}. Finally, we conclude in Sec.~\ref{sec:conclusions}.

\section{\label{sec:method} Evidence Calculations in Transformed and Marginalized Spaces of BNS mergers}

In this section, we outline methods to calculate the evidence of an equation of state model starting from an EOS agnostic \ac{PE} of \ac{NS} tidal deformability. The \ac{PE} data can be expressed in various equivalent spaces, with the computation of evidence in each representation offering distinct advantages.

\subsection{Bayesian Evidence Calculation using EOS-agnostic \ac{PE} results}

\par
Bayes Formula for posterior \ac{PDF} calculation for \ac{BNS} mergers can be written as
\begin{equation}
    \mathcal{L}(d_n|H_{GR},I) =  Z_{GR}\frac{p(\Theta|d_n,H_{GR},I)}{p(\Theta|H_{GR},I)}
    \label{eq:likelihoodGR}
\end{equation}
where, the prior assumed on mass and $\Lambda$ parameter ($p(m,\Lambda|H_{GR})$) contains the maximum and minimum values of parameter possible among all possible EOS hypotheses. $Z_{GR}$ is the marginalized likelihood also known as Bayesian evidence of the assumptions that general relativity is correct theory of gravity and the event is a binary neutron star merger without any assumptions on the EOS; we call this hypothesis $H_{GR}$. $Z_{GR}$ is calculated as 
\begin{equation}
    Z_{GR} = \int_{U(\Theta)} \mathcal{L}(d_n|H_{GR},I) p(\Theta|H_{GR}) d^n\Theta
    \label{eqn:def-evidence-GR}
\end{equation}
where $n$ is the dimensionality of the prior parameter space which in our case is 4 for the case of \ac{BNS} mergers. The total number of parameter for \ac{GW} inference of \ac{BNS} mergers has 17 parameters but, in our calculations, we start with the posterior \ac{PDF} which is marginalized along all parameters except two masses ($m_1,m_2$) and two tidal deformabilities ($\Lambda_1,\Lambda_2$).

To address the problem of \ac{NS} \ac{EOS}, we would like to calculate the evidence given the \ac{BNS} \ac{GW} data for a class of EOS hypotheses using the posterior distribution derived from an EOS agnostic \ac{PE} run. This is desirable to significantly reduce the computation time compared to conducting \ac{PE} runs for each EOS hypothesis separately. Using this approach, the evidence for any specific EOS hypothesis can be determined by performing an integration over the smaller relevant region of the EOS-agnostic range of the prior parameter space, adhering to the known relationship between the tidal deformability parameter $\Lambda$ and the mass $m$ of the NS corresponding to a given EOS. Formally, a set of EOSs can be viewed as a set of hypotheses $\{H_k\}$,
\begin{equation}
H_k: \theta_k(m_1,m_2)
\end{equation}
where $\theta_k$ are the tidal deformability parameters either $(\Lambda_1,\Lambda_2)$ or $(\tilde{\Lambda},\delta\tilde{\Lambda})$\footnote{Properties of NS uniquely dependent upon the mass can also be used in place of as well as in addition to quadrupolar tidal deformability such as higher order tidal deformability and spin-induced quadrupole moment.}. The likelihood function assuming an EOS can be written using delta function and the EOS-agnostic likelihood (assuming only general relativity) as follows:
\begin{equation}
\mathcal{L}(d_n|H_k,H_{GR},I)=  \delta(\Theta - \theta_k(m)) \mathcal{L}(d_n|H_{GR},I)  ~. \\
\label{eq:likelihood}
\end{equation}

where the likelihood on the right-hand side is posterior PDF divided by the EOS-agnostic prior on $(m,\Lambda)$ under the assumption, $H_{GR}$. Parameters $\theta_k(m)$ are a subset of parameters $\Theta$ that are fixed by the assumption of an EOS. We take $\theta_k(m)$ to be the tidal deformability with respect to the mass, \ie $\Lambda(m)$, given by the assumption $k$ of an EOS. Hence, the integration space of EOS-specific evidence calculation can be written as $\overline{\theta}=\Theta - \theta_k$ in the sense of mathematical sets.

The assumption of an EOS hypothesis ($H_k$) fixes the value of $\Lambda$ for a given mass along with limiting the upper limit on mass prior. Although the minimum mass possible for a particular EOS is quite uncertain, we take that to be 1 $M_\odot$ \citep{Suwa:2018uni} while the maximum is what is allowed in the stable branch of the \ac{TOV} sequence for that particular EOS.

The integration that yields the evidence calculation for a hypothesis $H_k$ will be performed over the range of prior space allowed by that particular EOS with the formula 

\begin{equation}
    Z_{k} = \int_{U(\overline{\theta})} {\cal L}(d_n|H_k,H_{GR}, I) p(\overline{\theta}|H_k,H_{GR},I) d^{m}\overline{\theta}
\label{eq:evidence-EOS}
\end{equation}
where the value of $m$ is 4 for the evidence calculation of sec~\ref{sec:4dEvidence} and 2 for calculations in sec~\ref{sec:2d-lamteta} and sec~\ref{sec:2d-mlam}.

Hence, using Eqn~\ref{eq:likelihoodGR} and Eq~\ref{eq:likelihood} in Eq~\ref{eq:evidence-EOS} the evidence for an EOS $Z_k$ would be  

\begin{equation}
\tilde{Z}_{k} \int_{U(\overline{\theta})} d^m\overline{\theta}  \delta(\Theta-\theta_k(m)) p(\Theta|d_n,H_{GR},I) \frac{p(\overline{\theta}|H_k,H_{GR},I)}{p(\Theta|H_{GR},I)}.
\label{eqn:model-evidence}
\end{equation}

Please note that $Z_{GR}$ in Eq~\ref{eqn:def-evidence-GR} is generally calculated over a larger region of parameter space than is allowed by any particular EOS for its evidence calculation, i.e. $\overline{\theta} \subset \Theta$. Therefore, evidence for an EOS will contain, apart from the prior ratio, an extra EOS-dependent suppression constant, which can be obtained from the posterior PDF of the GR. This will be greater for EOS, allowing for a smaller range of mass and tidal deformabilities. This extra factor combined with $Z_{GR}$ is written here as $\Tilde{Z}_{k}$ in Eq~\ref{eqn:model-evidence}.

In our analysis, we derive the prior and posterior density estimates utilizing prior and posterior samples correspondingly. For each space, the samples are first transformed to another space where the posterior density is then computed. This is akin to applying transformations with Jacobians and additional marginalization, especially when integration is carried out in a lower-dimensional space.
	
The factor $\tilde{Z}_k$ addresses the variance in effective prior volume between our analysis and the EOS-agnostic parameter estimation (PE) approaches. This difference emerges as our method enforces a narrow prior on the chirp mass $\mathcal{M}$ to adhere to the specifications of the relative binning method in \texttt{bilby}. Hence, the evidence integration range in our study is considerably smaller than that in the EOS-agnostic PE, except when the event is near the maximum mass stipulated by a specific EOS. As a result, $\tilde{Z}_k$ deviates noticeably from unity (typically $>0.97$) for only a limited number of events, generally less than 10\%, where the maximum allowable mass by the EOS is close to the observed posterior support. For the majority of events, this difference remains minimal. We have confirmed that this slight deviation of $\tilde{Z}_k$ from 1.0 holds no substantial influence on our findings, since it impacts only a small portion of the sample and does not affect the overall trends of the Bayes factor.


\begin{figure}[htb]
    \includegraphics[width=0.95\linewidth, trim = 0cm 0cm 0cm 0cm]{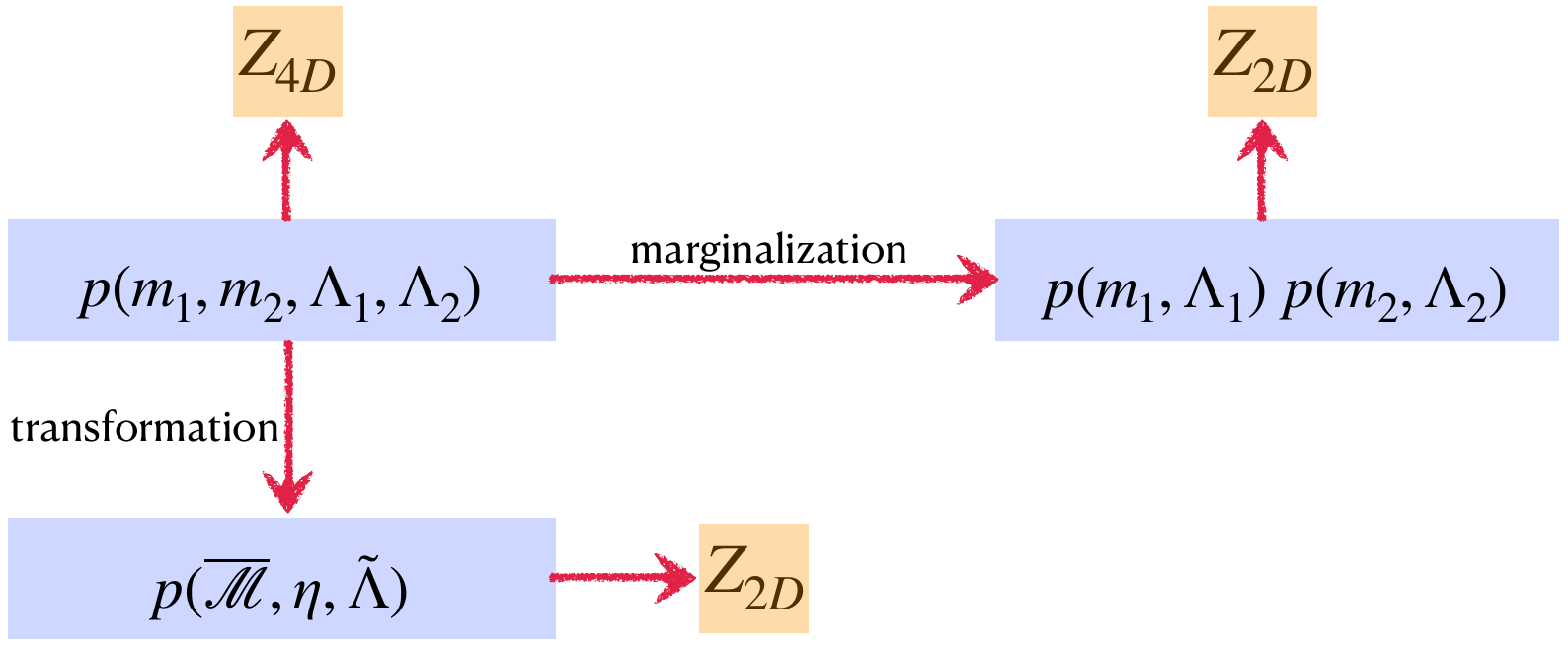}
    \caption{Here we show the flow chart of our calculation of evidence in different spaces and their relations as discussed in sec~\ref{sec:method}. We start with the marginalized 4D posterior, $(m_1,m_2,\Lambda_1,\Lambda_2)$ from the Bilby \citep{LALSuite2020} \ac{PE} results and marginalize it to obtain two 2D posterior in $m_1-\Lambda_1$ and $m_2-\Lambda_2$. The 3D posterior, $\mathcal{M},\eta,\tilde{\Lambda}$ is obtained by transforming the 4D posterior to $(\mathcal{M},\eta,\tilde{\Lambda},\delta\tilde{\Lambda})$ and ignoring $\delta\tilde{\Lambda}$ for computational efficiency. In some cases, it is also useful to take a mean value of $\mathcal{M}$ reducing the 3D posterior to just 2D, $(\eta,\tilde{\Lambda})$ with negligible loss of the evidence.}
    \label{fig:flowchart}
\end{figure}

In marginal cases where posterior samples, pertaining to component masses, exceed the maximum mass dictated by an EOS model, these samples are excised, and integration is conducted only up to this maximum mass. For events that occur in proximity to the maximum mass, the exclusion of samples beyond this threshold results in a negligible loss of posterior samples (within the integration range) and does not significantly impact the accuracy of our evidence calculation, assuming that EOS-agnostic posterior samples are appropriately normalized.

We next describe this general process for the problem of GW observations of the BNS inspiral in different equivalent spaces, as described as a flow chart in Fig.~\ref{fig:flowchart}.

\subsubsection{Evidence Calculation in 4D PDF, $p(m_1,\Lambda_1,m_2,\Lambda_2)$}\label{sec:4dEvidence}

We follow and expand upon the method originally derived in \citet{Bapna2018}, where the joint PDF $p(m_1,m_2,\Lambda_1,\Lambda_2|d)$ is used for the calculation of evidence. We drop $H_{GR}$ and $I$ from the PDF expressions as it will be implicitly assumed from here on. As shown in Eqn.~\ref{eqn:4DEvidence}, the two Dirac-delta functions signify the precisely known value of tidal deformability given masses from the posterior samples. In the case of poor EOS modeling and similar effects, one can consider a distribution of $\Lambda$ for a given mass, which may encode the uncertainty in the ab initio calculations of EOS models. We assume the same EOS model for both \ac{NS} components here.

\onecolumngrid

\begin{align}\label{eqn:4DEvidence}
Z_{k} &= \tilde{Z}_{k} \int dm_1 \int dm_2 \int d\Lambda_2 \int d\Lambda_1  \delta(\Lambda_1-\Lambda_k(m_1)) \delta(\Lambda_2-\Lambda_k(m_2)) P(m_1,\Lambda_1,m_2,\Lambda_2|d_n) \frac{P(m_1,m_2|H_k,H_{GR},I)}{P(m_1,m_2,\Lambda_1,\Lambda_2|H_{GR},I)}
\end{align}

\twocolumngrid

\subsubsection{Evidence calculation in the $\Tilde{\Lambda}-\eta$ space}
\label{sec:2d-lamteta}

The leading term affecting the GW emission from a BNS system arises at fifth \ac{PN} order beyond the dominant quadrupole term \cite{Flanagan:2007ix, Hinderer:2007mb, Hinderer:2009ca, Vines:2011ud}, where a certain combination of individual quadrupolar tidal deformabilities called effective tidal deformability, $\tilde{\Lambda}$ appear. To compare different methods, we transform the posterior distribution to $\Tilde{\Lambda}$-$\delta\Tilde{\Lambda}$ space. As highlighted in the literature \citep{Ghosh:2021eqv}, the posterior distribution of $\delta\Tilde{\Lambda}$ is usually noninformative. Hence, following \citealt{Ghosh:2021eqv}, we perform an evidence calculation in the $\mathcal{M},\eta,\Tilde{\Lambda}$ space fixing $\mathcal{M}$ to its median value $\overline{\mathcal{M}}$, since the chirp mass is a much better measured parameter compared to $\eta$ and $\Tilde{\Lambda}$.\\

In the 2D space the evidence for a given EOS is 
\begin{align}
& Z_{k}  = \int d \eta \int d \tilde{\Lambda} \delta\left(\tilde{\Lambda}-\tilde{\Lambda}_k(\overline{\mathcal{M}}, \eta)\right) p(\overline{\mathcal{M}}, \eta, \tilde{\Lambda}|d_n). 
\label{eqn:evi-mcetalamt}
\end{align}

\subsubsection{Evidence calculation in the subsapces $m_1-\Lambda_1$ and $m_2-\Lambda_2$}
\label{sec:2d-mlam}

We propose another way to calculate the evidence directly in $m$-$\Lambda$ space. To this end we assume the pair $(m_1,\Lambda_1)$ to be statistically independent of the pair $(m_2,\Lambda_2)$. With this assumption, the 4D PDF $p(m_1,\Lambda_1,m_2\Lambda_2)$ can be expressed as a product of individual PDFs of mass-tidal deformability,
$$
p(m_1,\Lambda_1) = \int \int p(m_1,\Lambda_1,m_2,\Lambda_2|d_n) dm_2 d\Lambda_2
$$
and similarly for $p(m_2,\Lambda_2)$.
 
The evidence for an EOS using PDFs from both companions is 
\begin{align}
     Z_{k,i}=\int_{U(\Theta_{k})} p(m_i,\Lambda_i|d_n) \delta(\Lambda_i-\Lambda_k(m)) dm_i d\Lambda_i  \quad i=1,2 \label{eqn:z-singleNS} 
\end{align}
using which we calculate the evidence of an EOS for single BNS events as product of two evidences, $Z_k = Z_{k,1} \times Z_{k,2}$. In this procedure, while we obtain the marginalized pdf by integrating out the information from the companion star, multiplication of evidence implies that we take the 4D PDF as product of two 2D PDFs. We ignore the correlation between masses which is justified because it does not contain the information about the EOS but only about the mass-ratio property of the population of the BNS events occurring in the observable universe.

\subsection{Combining Evidence from multiple events}\label{subsec:CombiningZn}

The number of BNS events with high significance (SNR $>10$) is expected to increase from a couple in the O5 observing runs of LVK \citep{Kiendrebeogo:2023hzf} to tens of thousands in a planned network of GW detectors with Einstein Telescope \citep{Hild2008, Hild2010, Hild2011} and two Cosmic Explorers \citep{Evans2023-wh,Gupta:2023lga,Pandey:2024mlo}. For the case of model selection, the relevant quantity is the odds ratio given by:
\begin{equation}
[O^{i}_{j}]_n=\frac{Z(H_i|d_n,H_{GR},I)}{Z(H_j|d_n,H_{GR},I)} 
\end{equation}
Here, $H_k$ is the EoS hypothesis that gives $\Theta = \Theta_k (m)$ as described above. 

Evidence from multiple events can be combined to obtain a cumulative odds ratio for better results. Considering equal prior probabilities for all EOS and all events being independent, the cumulative odds ratio (on logarithmic scale) is,

\onecolumngrid

\begin{eqnarray}
\log \sum [O^i_j]_n &=& k \log f_{PC} + \log \bigg[ \frac{Z(H_i|d_1,d_2,....,d_N,I)}{Z(H_j|d_1,d_2,....,d_N,I)}\bigg] \\
&=& k \log f_{PC} + \sum_{n=1}^{N}( \log[Z(H_i|d_n,I)]-\log[Z(H_j|d_n,I)] ) \nonumber
\end{eqnarray}
\twocolumngrid

where $f_{PC}=(M_{max,j}-M_{low})/(M_{max,i}-M_{low})$ is the EOS-specific component mass prior correction factor taken into account differently for each EOS. We take this to be a uniform distribution between $M_{low}=1 M_\odot$ and the maximum mass allowed by that specific EOS. The factor $k$ is there to undo the prior used for the calculation of evidence for each event with the following value: $k=N-1$ for the calculation of 3D and 4D evidence and $k=2N-1$ for the calculation of 2D evidence in the subspace. We keep our integration range slightly outside the range used in Bayesian PE run (except for the samples beyond the maximum mass of EOS), to allow for the \ac{KDE} to keep a conservative view of EOS-specific prior correction factor. We compared the \ac{KDE} with the underlying sample distribution and found an excellent agreement (see Fig.\ref{fig:kde_scatter}), effectively ruling out the risk of errors arising from the choice of density estimation method.

\subsection{Injections and \ac{PE} Runs using Relative Binning Framework in Bilby}

\subsubsection{Choice of EOS}
\label{subsubsec:EOS-choice}

To establish the robustness of our model selection method, we selected a collection of EOSs characterized by diverse compositions and formulations. The nucleonic EOSs we consider are: DD2 \cite{Typel:2009sy, Hempel:2009mc}, LS220 \cite{Lattimer:1991nc}, SFHo \cite{Steiner:2012rk}, APR3, APR4 \citep{Akmal:1998cf}, and Sly \citep{Douchin:2001sv}. Furthermore, we consider two EOS models that include hyperons in addition to nucleons: BHB \cite{Banik:2014qja}, H4 \cite{Glendenning:1991es, Lackey:2005tk, Read:2008iy}. Additionally, the ALF2 \cite{Alford:2004pf} EOS incorporates a phase transition to deconfined quarks and PP2 and PP5 are two piecewise polytropic EOSs taken randomly from 2 million such samples provided by \citet{Godzieba:2020bbz}. In choosing EOSs, our primary objective is to develop a method that can robustly and efficiently encapsulate the intricacies of the EOS models while adhering to the principle of Occam's razor.\\

Formally, to compare two models, it is essential to measure the difference between them, which has traditionally been referred to by the names of EOSs, highlighting the fundamental assumptions in the construction of EOSs. Although such concepts of distinguishability have significance, they are not useful in quantifying their differences. We formally quantify the disparity between different EOSs as the $L_2$ distance between them (refer to Eqn. (14) of \citet{Kashyap:2022wzr} and ~\ref{apdx:L2}), with the caveat that this metric does not entirely capture their differences. In addition, other distance measures and correlations with populations will be crucial considerations, which we aim to explore in our upcoming research.

\begin{figure}[htb]
    \centering
    \includegraphics[width=1.0\linewidth, trim = 0cm 1cm 0cm 2cm]{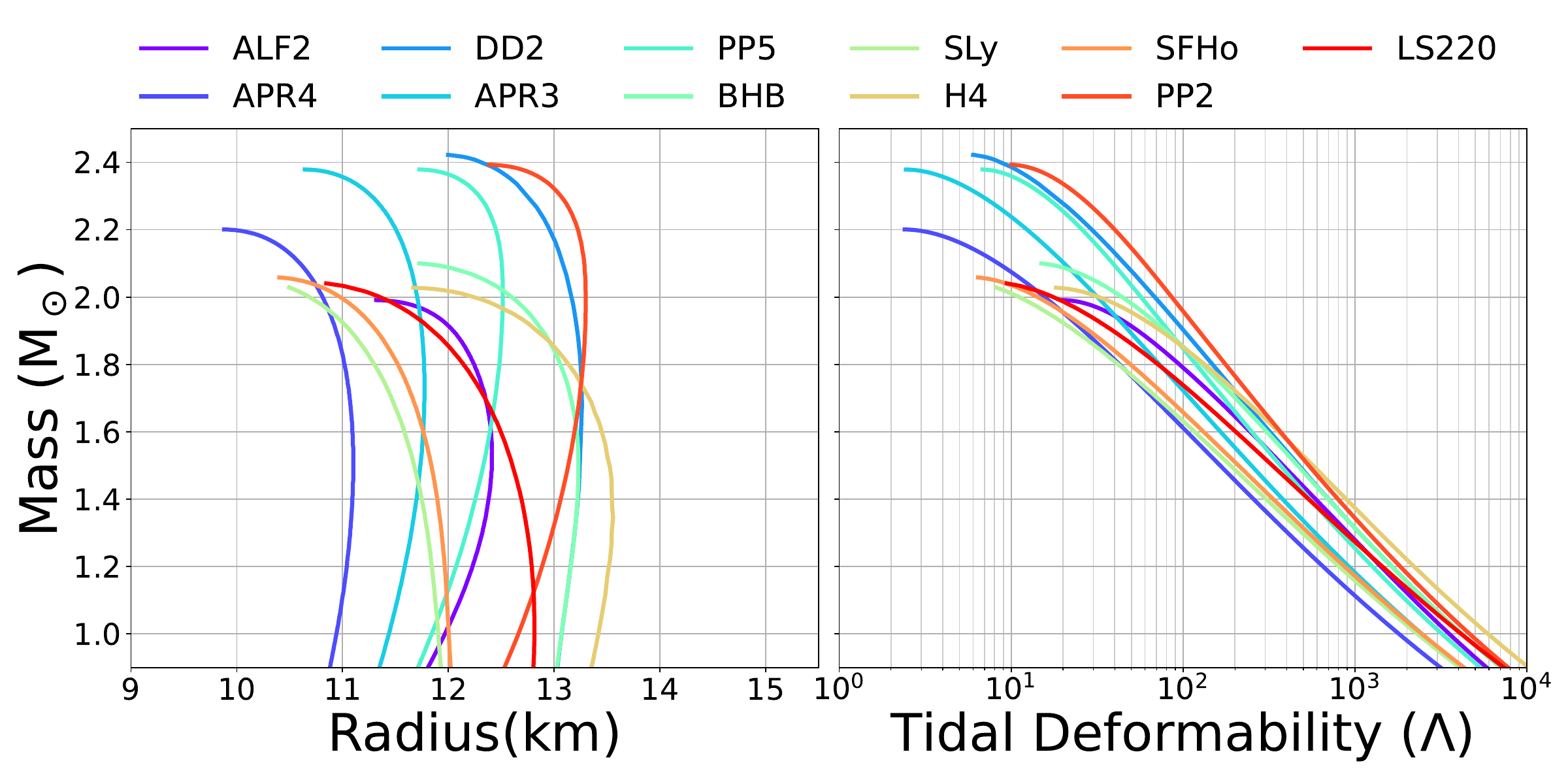}
    \caption{Mass-Radius-Tidal deformability curves for our choice of EOSs used in the current paper. We have chosen EOSs varying in stiffness, compactness, maximum mass and the formalism used for construction. EOS surves span wide variety of physical phenomena in dense matter in order to test the robustness of the method presented here.}
    \label{fig:mr-curve}
\end{figure}

In our model selection method, we strategically refrain from mapping the data to the $p$-$\epsilon$ space, which significantly influences parameterization, primarily phenomenologically \citep{Legred:2022pyp, Greif:2018njt}. These phenomenological parameterizations introduce artificial correlations for which we refer the reader to Refs.~\citep{Legred:2022pyp, Greif:2018njt}. For the most recent Bayesian model selection outcomes that include mapping to the pressure-density curve, we direct the reader to Refs.~\citep{Abdelsalhin2018-yb, Pacilio:2021jmq}. 

\subsubsection{\ac{PE} using Relative Binning}
We conduct Bayesian \ac{PE} on a simulated ensemble of binary neutron star (BNS) mergers, utilizing solely the gravitational waveform emanating from the inspiral phase of the merger. The parameters of the simulated BNS systems follow astrophysically motivated distributions for component masses and spins. The component masses are selected from a double Gaussian distribution, with the first Gaussian parameterized by mean $\mu = 1.35\,M_{\odot}$ and spread $\sigma = 0.08\,M_{\odot}$ and the second Gaussian by $\mu = 1.8\,M_{\odot}$ and spread $\sigma = 0.3\,M_{\odot}$ \cite{Gupta:2023lga}. We selected three EOSs and set the tidal deformability of the injected inspiral GW signals corresponding to the mass of the component NSs. The maximum mass of NS is fixed to the maximum mass allowed by the corresponding EOS. For NS spins, the astrophysical distribution of pulsar periods and assumptions about spin-down rates lead to birth periods in the range $10$-$140 \mathrm{~ms}$ \citep{Ott:2005wh}, which corresponds to dimensionless spins $\chi \equiv J / \mathrm{m}^2 \lesssim 0.04$ (see \citet{Gupta:2024bqn} for a discussion on NS spins). Furthermore, the fastest known pulsar in a BNS system has $\chi \sim 0.02$ \citep{manchester2017millisecond}. For the simulated population, we conservatively assume that the spins of merging \acp{NS} are small and aligned with the orbital angular momentum, but non-negligible for the analysis, with individual dimensionless spins sampled from a uniform distribution between $[-0.05,0.05]$.

The GW signals corresponding to these sources are generated using the \texttt{IMRPhenomPv2\_NRTidalv2} waveform \cite{Hannam:2013oca, Khan:2018fmp, Dietrich:2019kaq}. This spin-precessing frequency domain waveform model with the quadrupolar $(2,2)$ mode contains the dominant tidal contribution to the phase of the GW waveform at the 5PN, as well as the corrections to the phase evolution at lower \ac{PN} orders due to nonzero spin-induced quadrupole moments of \acp{NS}. The sources are uniformly distributed in comoving volume up to a distance of $450\,\mathrm{Mpc}$, equivalent to a redshift of $z\approx0.1$. With a local BNS merger rate of $320\,{\rm Gpc}^{-3}\,{\rm yr}^{-1}$ \citep{Abbott2021ApJPop}, we expect $\sim100$ BNS mergers every year within this distance \cite{Gupta:2023lga}.
The sky location $(\theta, \phi)$, inclination angle $(\iota)$, and polarization angle $(\psi)$ are all uniformly distributed on a sphere. The phase at the coalescence $(\varphi_c)$ is fixed to be $0$ rad. 

We perform Bayesian \ac{PE} on the BNS events in zero noise to obtain posterior distributions on binary parameters. This is accomplished using the implementation of relative binning \citep{Cornish:2010kf, Zackay:2018qdy, Krishna:2023bug} in Bilby \cite{2019ApJS..241...27A, 2020MNRAS.499.3295R}. The relative binning technique approximates the Bayesian likelihood as a piecewise linear function in the frequency domain, improving the likelihood evaluation speed without significantly compromising accuracy. 

As the likelihood is extremely sensitive to the chirp mass, the accuracy of the method is compromised when the likelihood is evaluated far away from the maximum likelihood value of the chirp mass (which, for zero-noise injections, is the injected chirp mass). Thus, while relative binning requires stringent bounds on the chirp mass in the prior space, these bounds are much broader than the inferred posterior distribution and do not bias the \ac{PE}. For Bayesian analysis, the mass priors are chosen to be uniform in component masses but constrained within lower and upper bounds on the chirp mass and mass ratio. Spin parameters follow aligned spin priors between spin magnitude of [0, 0.1] and the prior on luminosity distance is uniform in comoving volume between [10, 800] Mpc. We sample uniformly over the tidal deformability parameters, $\tilde{\Lambda}$ and $\delta\tilde{\Lambda}$ while the prior individual tidal deformabilities, $\Lambda_1$ and $\Lambda_2$ are inherited from $\tilde{\Lambda}$ and $\delta\tilde{\Lambda}$ priors and constrained between [0, 5000].

\section{\label{sec:Results}Results}

\subsection{\label{sec:bilbyruns} \ac{PE} Runs}

We perform Bayesian PE using the relative binning framework within the Bilby pipeline \citep{krishna2023bilbyrelbinpaper}, for 100 BNS events for each choice of EOS. The tidal deformability values for injected waveforms are taken for three EOSs: APR4, DD2, and SLy. Although the waveform model used (\texttt{IMRPhenomNRTidalv2}) has only a single combination of the two tidal deformabilities at the fifth \ac{PN} order \ie, that is, $\tilde{\Lambda}$, we sample in $\Lambda_1$ and $\Lambda_2$ independently while using the spin-induced quadrupole moment to break the degeneracy between two tidal deformabilities \citep{Dietrich:2019kaq}. Hence, we get independent constraints on the tidal deformability of both components.

\begin{figure}[tbp]
    \includegraphics[width=1.03\linewidth]{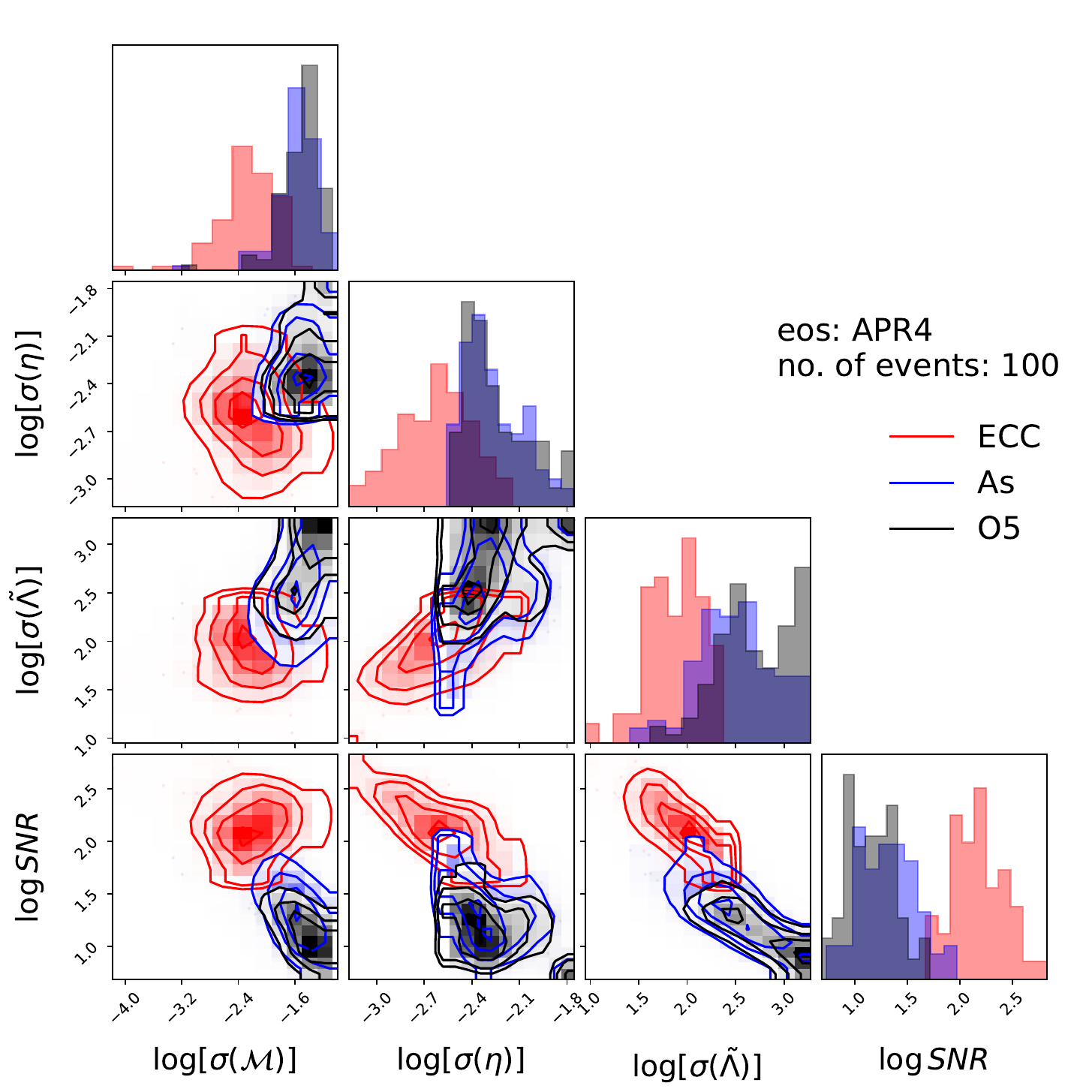}
    \caption{In this figure, we show the expected distributions and improvements in the measurement of three key quantities relevant for neutron star EOS problem. We plot the 1$\sigma$ full-width (on $\log_{10}$ scale) of the posteriors obtained from the \ac{PE} runs of 100 events where injected EOS is APR4. The A$\#$ will bring only slight improvement in the measurement of $\mathcal{M}$, $\eta$ and $\tilde{\Lambda}$ while the proposed ECC configuration is expected to increase the accuracy of the measurement by 1-2 orders of magnitude or higher compared to our current accuracy. As the detector network sensitivity increases we also expect to see few events with exceptional measurement leading to golden binaries that may allow us to perform precision of tests of gravity. The runs with other EOSs have similar distributions where exactly same set of parameters have been used above.}
    \label{fig:distEvidence}
\end{figure}
Figure~\ref{fig:distEvidence} illustrates the error distribution for all events across various detector configurations. The O5 network incorporates the LIGO observatories located in Hanford and Livingston \cite{LIGOScientific:2014pky,aLIGO:2020wna,Tse:2019wcy}, the Virgo detector in Italy \cite{VIRGO:2014yos,Virgo:2019juy}, and the KAGRA detector in Japan \cite{Somiya:2011np,KAGRA:2020agh,Aso:2013eba}, all operating at A+ or comparable sensitivities \cite{Miller:2014kma,KAGRA:2013rdx}. With this upgrade, the LIGO detectors are projected to achieve approximately 50\% greater sensitivity compared to their advanced LIGO counterparts. The \As network (denoted as ``As" in the figure) consists of three LIGO detectors located in Hanford, Livingston, and Aundh (India) \cite{LIGO-India}, operating at \As sensitivity. This configuration incorporates heavier test masses (100 kg) and a higher-power laser, with an anticipated operational start date of 2029 \citep{fritschel2022report}. Designed as an intermediate step between current detectors and future advanced configurations, the \As network bridges the gap towards the most sophisticated planned setups, represented by the ECC network. The ECC network includes the Einstein Telescope in Europe \cite{Punturo:2010zz,Hild:2010id,Branchesi:2023mws} and two Cosmic Explorer detectors with arm lengths of 20 km and 40 km, respectively \cite{Evans:2021gyd,LIGOScientific:2016wof,Reitze:2019iox}, in the United States. The increased scale and technological advancements of these detectors provide sensitivity improvements of $\mathcal{O}(10)-\mathcal{O}(100)$, depending on the frequency, relative to A+.

At a detection threshold SNR of 10, the O5 network identifies 70 out of 100 binary neutron star (BNS) mergers, the \As network detects 88 events, and the ECC network detects all 100 events. Furthermore, the \As network achieves higher SNR for detections, enabling more precise parameter estimation for binaries compared to O5. The ECC network exhibits $\sim100$-fold higher SNR compared to O5, facilitating a 100-fold enhancement in the precision of chirp mass measurements.

\subsection{Distribution of evidence}

We calculate the evidence using the three methods described above and plot their distributions for 100 events in three columns of Fig. ~\ref{fig:distEvidence-DD2}, ~\ref{fig:distEvidence-APR4}, and ~\ref{fig:distEvidence-SLy} for EOS DD2, APR4 and SLy, respectively. In principle, for zero-noise, the injected EOS should have their mean evidences to be highest among all model EOSs in this framework. However, we expect that the presence of realistic noise changes the expected behavior of the evidence distribution due to several factors. For example, due to the low SNR of some of the events, the posterior will be broad, leading to low evidence values even for correct \acp{EOS}. In some cases, even for high SNR events, but with possible systematic biases, a wrong EOS might have greater evidence compared to the injected EOS. Such scenarios also arise when the population of events is chosen in some specific manner in its parameter space. Hence, the evidence distribution of a model encodes both systematic and statistical errors in a single description.

As anticipated, with regard to near-future detectors, the distributions of evidence have equally large variances for all EOS models, irrespective of injected EOSs. This result indicates, as will be explained in the subsequent sections, that for the O5 and \As detectors, a larger number of events are necessary to accurately infer the injected EOS. Additionally, it is found that for more sensitive future detectors, the evidence distributions, and especially the peak values, are higher for the injected EOS compared to other EOSs. This can be attributed to the increase in the SNR and the extended horizon distance of future detectors.

We observe the expected trend that nearby EOSs (check Fig.~\ref{fig:mr-curve}) exhibit similar evidence distributions as evident in Fig.~\ref{fig:distEvidence-DD2}. Specifically, for the ECC detectors and DD2 injection runs, the DD2 and BHB EOSs yield comparable evidence values in the $\tilde{\Lambda}$-$\eta$ space \citep{Ghosh:2021eqv}, while in $m$-$\Lambda$ space DD2 has the highest evidence across all EOSs. However, the evidence distribution for BHB is wider than that for DD2, demonstrating the robustness of the above two methods. Pairs of EOSs with closely matched evidence include APR4-SLy (for APR4 injection) and SLy-SFHo-APR4 (for SLy injection). We show these cases in Fig.~\ref{fig:distEvidence-APR4} and ~\ref{fig:distEvidence-SLy}, respectively. We stress that degeneracies can skew the determination of the correct \ac{EOS} model even when the most sensitive detectors are used. Consequently, any estimator of the statistical evidence for an EOS model must be thoroughly evaluated for accuracy and optimality.


\if\mycmd1


\begin{figure*}[pt]
    \centering
    \includegraphics[width=0.99\textwidth]{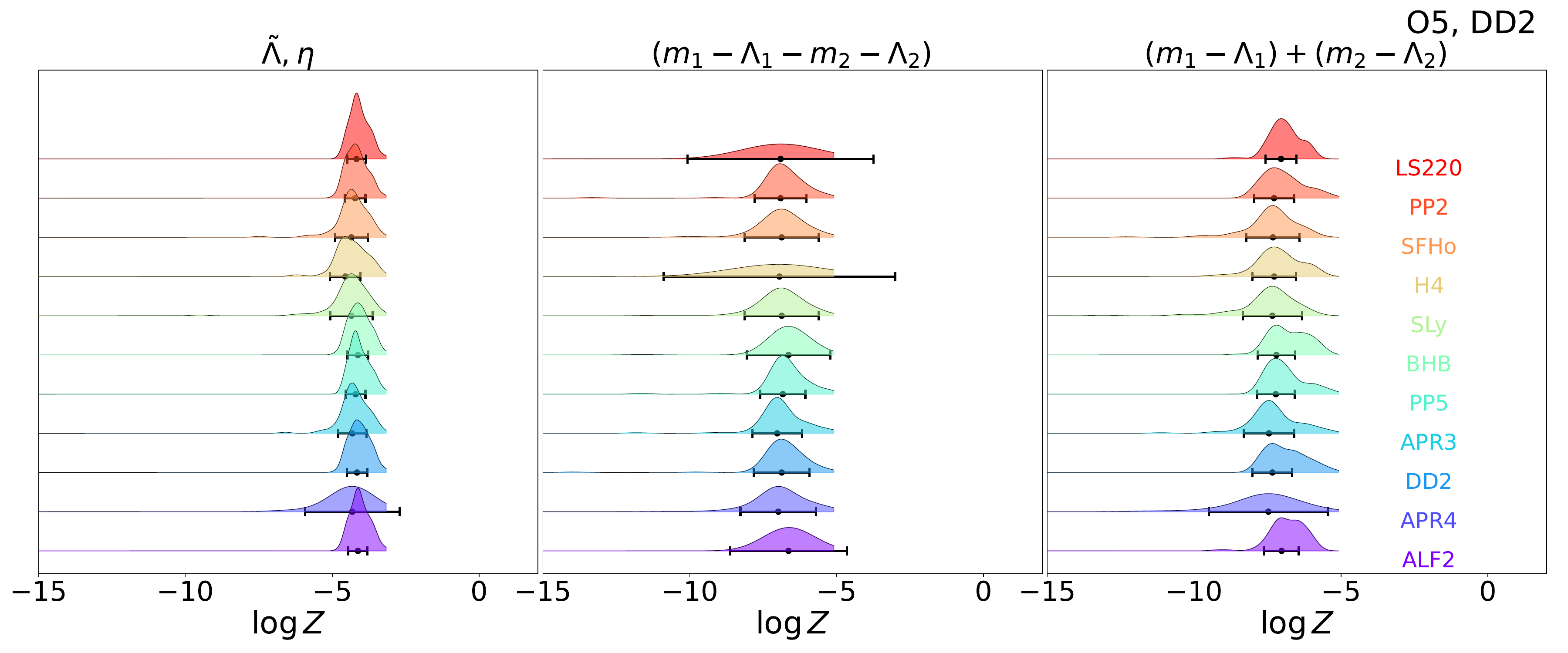}
    \includegraphics[width=0.99\textwidth]{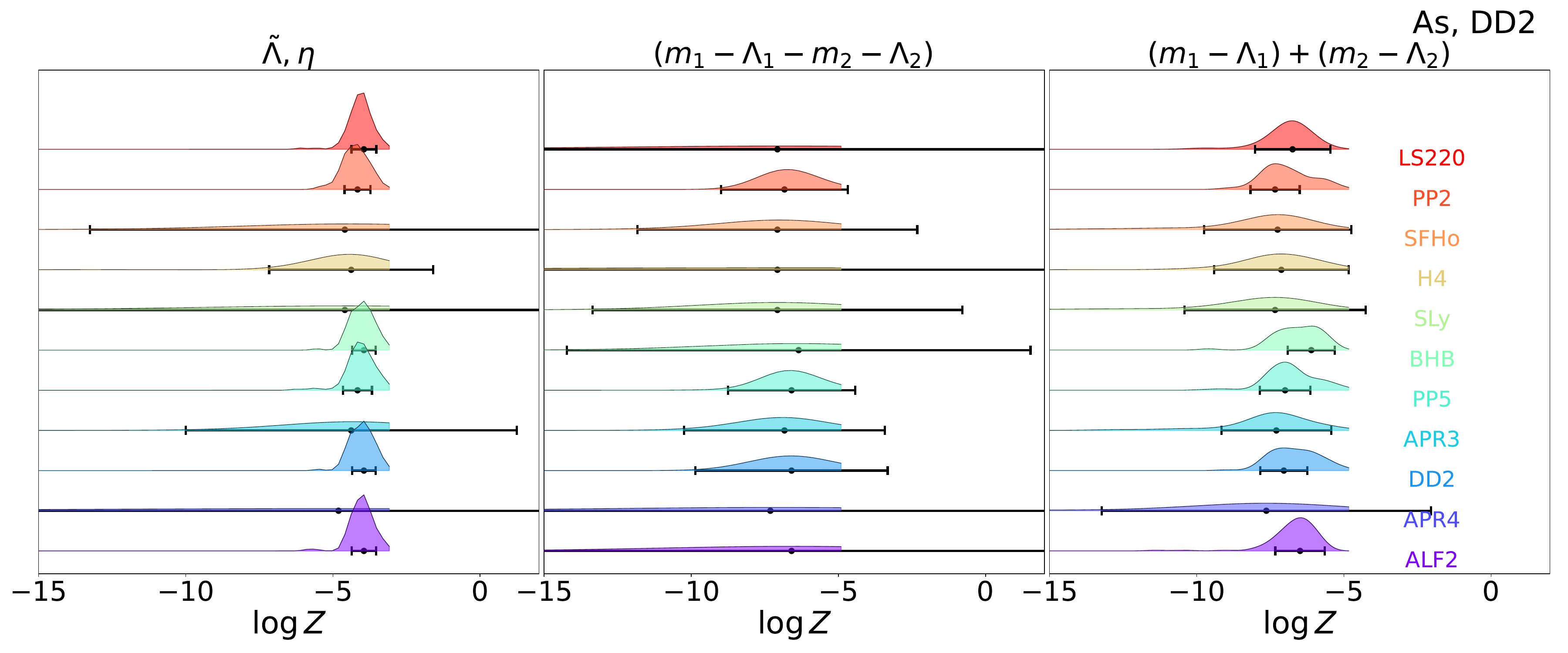}    
    \includegraphics[width=0.99\textwidth]{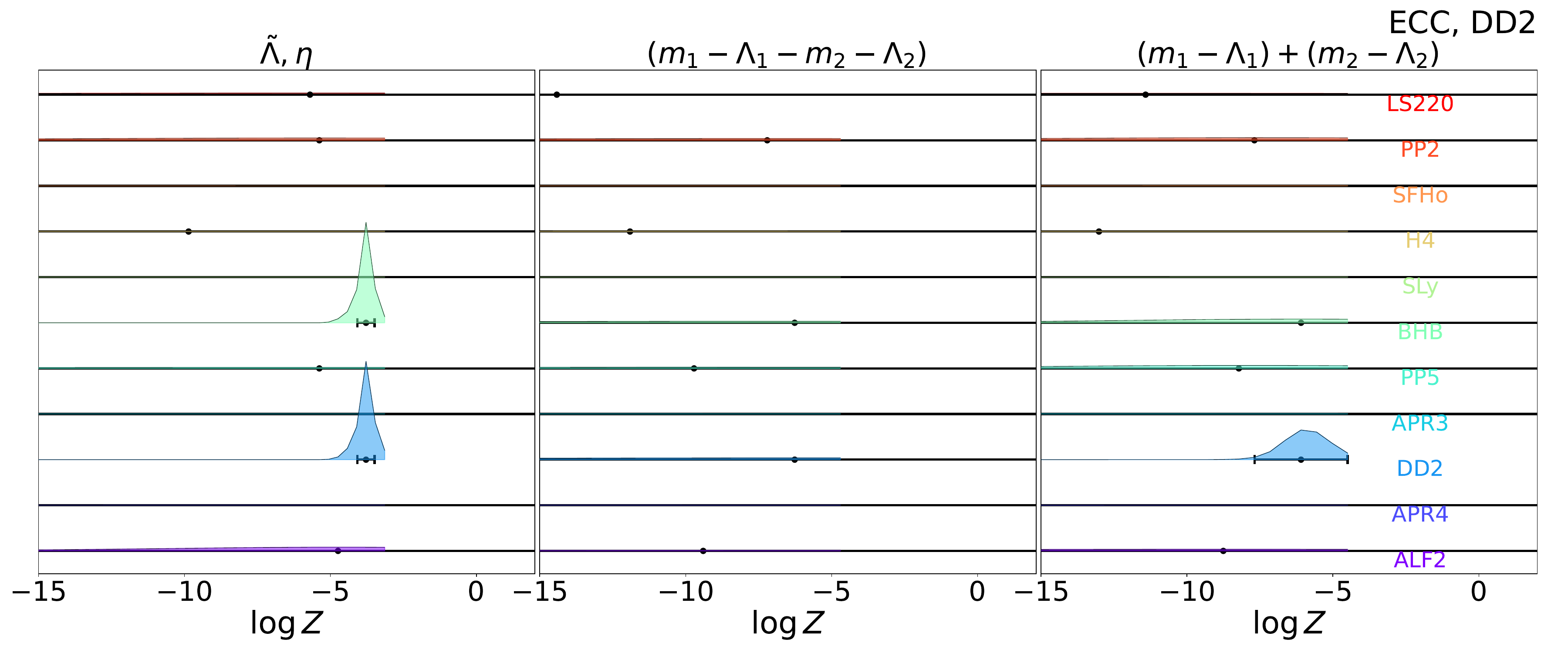}
    \caption{Here we show the distribution of evidence using the three methods described in section ~\ref{sec:method} and for our collection of 11 EOSs considered candidate models. Each ridgeline plot corresponding to a model is smoothened distribution of evidence of 100 events. The three panels show the results for three detector configurations where the injected EOS is DD2. We represent the median value and 1-$\sigma$ width with a filled circle and an error bar on the line, respectively. It is evident from the top two panels that \As (As in plot legend) does not bring much improvement in disinguishing EOSs while the ECC clearly picks up the correct EOS where the evidence for the DD2 EOS is large for most of the events. Please also note that errors reduce from the present to future detector configurations (top to bottom). The calculations in the subspace are highlighted by both distinguishing properties -- the median values being higher for the injected EOS and the width of the distribution being small, while the 4D space calculations are not informative even for our most promising detector confifurations. Similar plots for other EOS injection show similar behaviors with respect to detector networks and the model EOSs shown here.}
    \label{fig:distEvidence-DD2}
\end{figure*}


\fi

Degeneracies are anticipated to represent a fundamental limitation in differentiating distinct EOS models, as two different $m$-$\Lambda$ relationships may yield identical observed $(\tilde{\Lambda},\eta)$ posterior data. We find that the ambiguity in model selection due to the broad posterior distributions (due to the large noise in O5 and A$\#$) is greater than the apparent problem of degeneracy described above. Hence, we expect that $Z_{\tilde{\Lambda},\eta}$ will not be able to distinguish models in future GW detectors (such as ECC). However, we find that such differences are negligible and are important only when we combine the evidence from several events. As seen in Fig.~\ref{fig:BF-ECC}, the degeneracy in $Z_{\tilde{\Lambda},\eta}$ is shown where two EOSs compete with the injected EOS, SLy in this case, up to a large number of events. However, the subspace evidence $Z_{m-\Lambda}$ does slightly better.

\subsection{Cumulative Evidence from multiple BNS events}

Calculating the evidence for each event individually is not sufficient for the selection of the right EOS model simply because single \ac{BNS} provides information on the tidal deformability of only two points (corresponding to two component masses) on the EOS curve. With a population of events it should be possible to sample the \ac{NS} mass and tidal deformability over a wider range of priors. Hence, we calculate the joint evidence of an EOS model from multiple events. We plot this joint evidence as a function of the number of events, as described in sec.~\ref{sec:4dEvidence}. 

\begin{figure}[htb]
    \centering
    \includegraphics[width=0.45\textwidth]{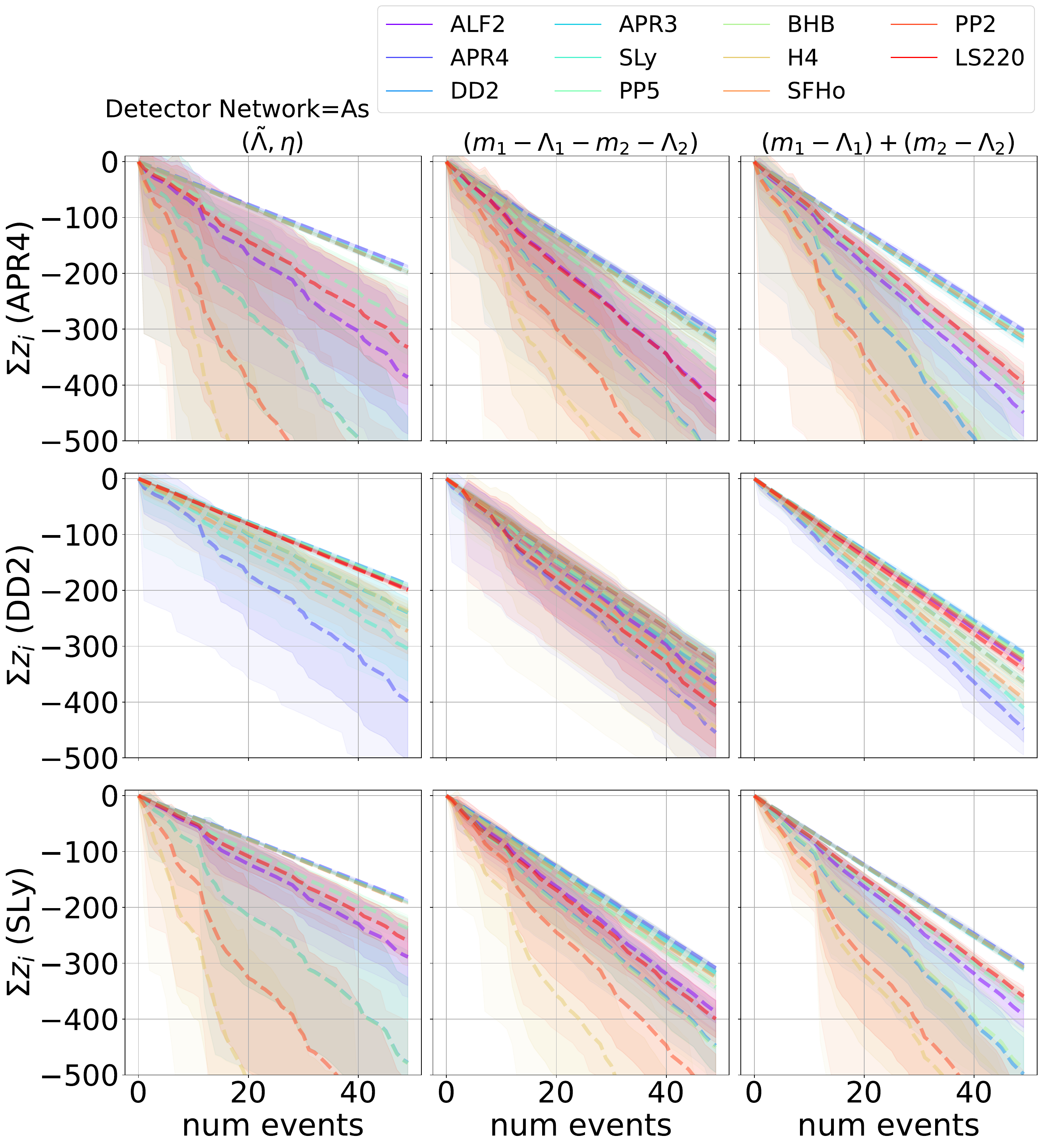}
    \caption{Here we plot the cumulative evidence on log scale of various EOS models as a function of the number of events with the three injection EOSs for detectors with \As sensitivity. The model with the highest evidence is the preferred model however confusion may arise based on the type of events selected. We show the uncertainties as color bands arising from the sampling errors from the set of events.}
    \label{fig:logevi-As}
\end{figure}

From a detailed analysis of a complete set of 100 events, we determine the cumulative evidence and Bayes factor for each model. The cumulative evidence is depicted as a function of the number of events in Figure~\ref{fig:logevi-As}. By repeatedly sampling a predefined number of events, we assess and quantify the statistical uncertainty inherent in the cumulative evidence calculations. This uncertainty is visually represented by a shaded region surrounding the median line in the figure. For all of the injected EOSs and detector networks, our methods give highest value of the evidence to the injected model asymptotically i.e. for a large number of events. We also see a reduction of the errors in the evidence of the injected EOS as the number of BNS events increases.

Additional clarity on detector sensitivity can be deduced from cumulative evidence plots. For example, the \As detector configuration is more capable of distinguishing a cluster of EOS that are similar to each other but significantly different from the injected EOS compared to O5 (Fig. ~\ref{fig:logevi-As}). In the case of optimistic ECC observatories, the evidence for some of the (noninjected) EOS models are numerically zero, as shown in Fig.~\ref{fig:L2vsBF-all}.

\subsubsection{Properties of Cumulative Bayes Factor Matrix}

\begin{figure}[htb]
    \centering
    \includegraphics[width=0.45\textwidth]{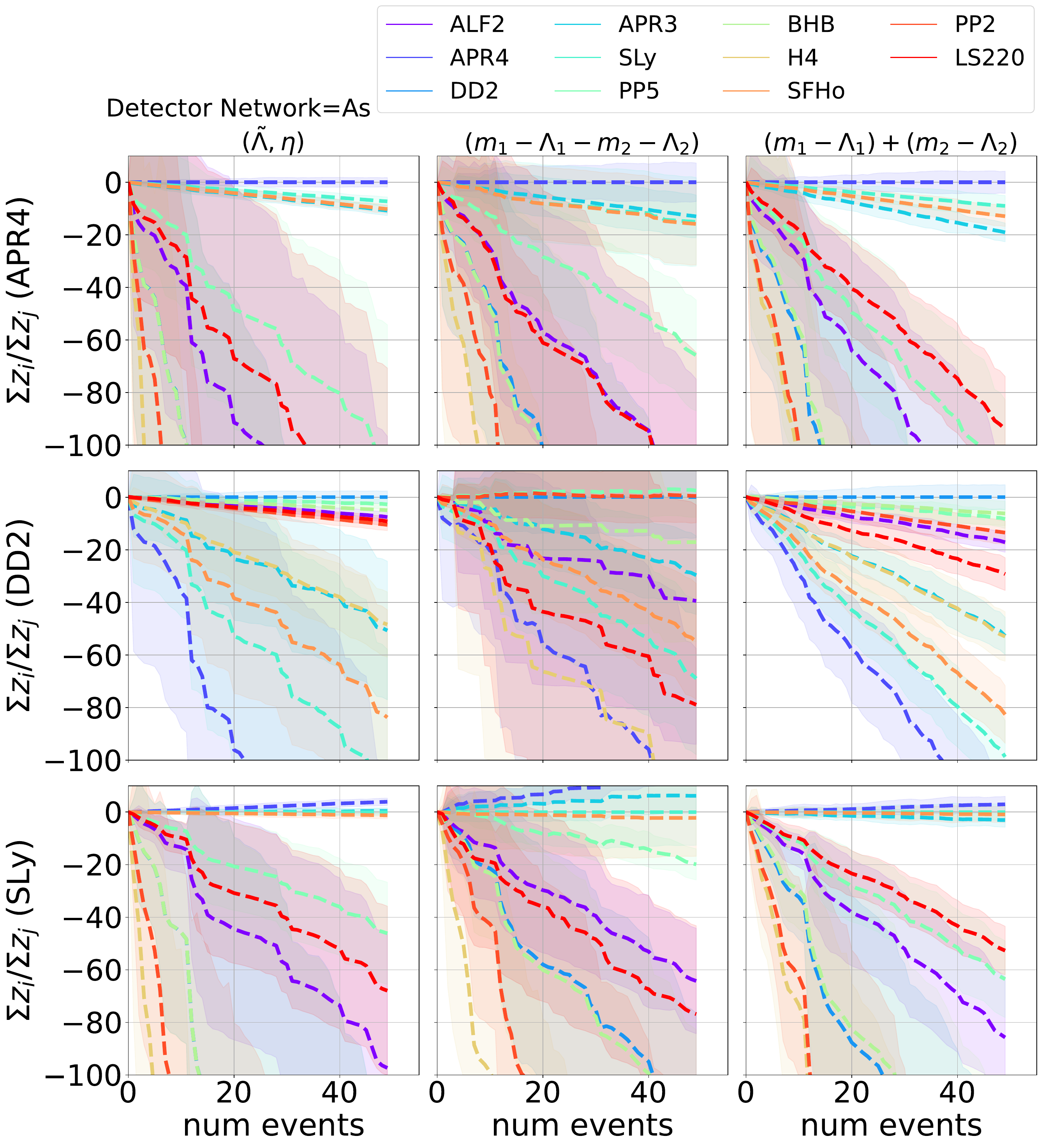}
    \caption{The Bayes Factor of 11 EOS models against the injected EOS in the population. The data from three injected EOSs are plotted in three rows, APR4, DD2 and SLy, respectively.  Negative values suggest that the corresponding EOS model (marked by different colors) is disfavored compared to the injected EOS. The Bayes Factor should be equal to zero for the injected EOS in our convention. Note that DD2 and BHB overlap for a large region of $m-\Lambda$ curve (c.f. Fig.~\ref{fig:mr-curve}) which is reflected in the evidence calculation as BHB continues to gain similar Bayes Factor values against DD2 even with large number of events.}
    \label{fig:BF-As}
\end{figure}

\begin{figure}[htb]
    \centering
    \includegraphics[width=0.45\textwidth]{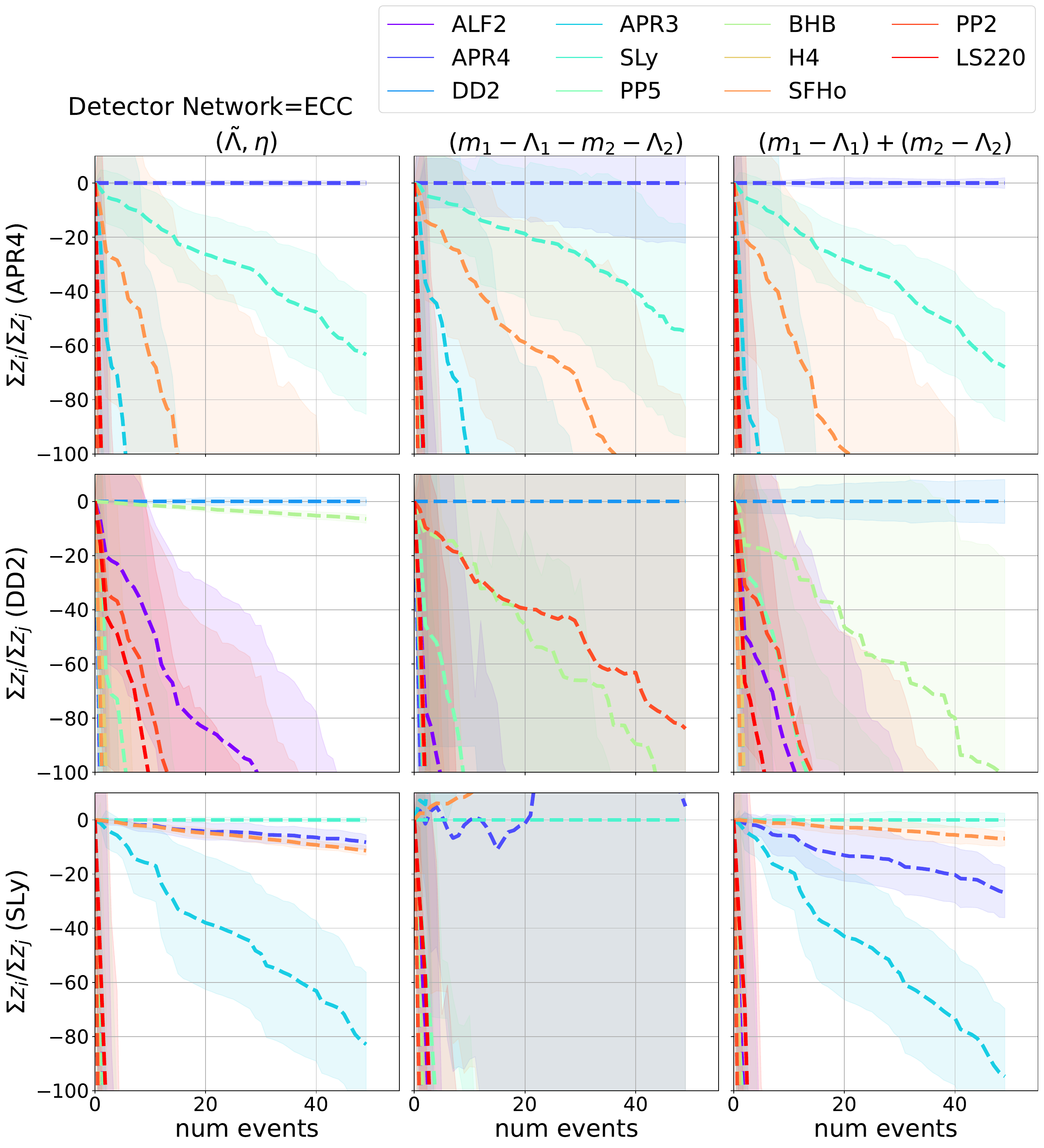} 
    \caption{same as Fig.~\ref{fig:BF-As} but for ECC configurations of the next-generation detectors with three injected EOSs as a function of the number of events.}
    \label{fig:BF-ECC}
\end{figure}

Given a distribution of evidence in different spaces, one can ask the joint probability of any EOS (mass-tidal deformability curve) against another EOS, i.e., the Bayes factor. This is usually represented as a confusion matrix where the diagonal terms represent the null hypothesis that the data contain the assumed model, while off-diagonal elements represent the Bayes factor between any two models from the collection. This matrix can be added for multiple independent events. For a robust model selection method, the diagonal elements of the cumulative matrix corresponding to the injected EOS should grow to be the maximum among all values in that particular row as the number of events accumulates. However, comparing two noninjected EOSs can lead to an unphysically large Bayes factor since both may gather very small evidences yet their ratio could be quite large. In this case, we can think of the model selection process as a two-step process in which we first select the EOS model with the largest cumulative evidence (derived from Fig.~\ref{fig:logevi-As}) and then calculate its Bayes factor against other models lower in the rank of total absolute evidence in a given space (as done in Fig.~\ref{fig:BF-As}). 

Figures ~\ref{fig:logevi-As} and ~\ref{fig:BF-As} show that with a sufficiently large number of events, the correct EOS has the highest cumulative evidence, as well as positive Bayes factor against the model with the next highest evidence. The cumulative log evidence plot can be used as the model ranking statistic. The number of events required to achieve this level of distinguishability varies for different detectors. As expected, it is smallest for ECC  network. 

\subsection{Criterion for distinguishability and Method Efficacy}

In realistic scenarios, we are tasked with navigating through a collection of models from which we need to find the model that best represents the data according to a statistical procedure. In addition, models are continuously distributed in the function space, $p$-$\epsilon$ (equivalently, $m$-$R$ or $m$-$\Lambda)$. The selection of the optimal statistical procedure is also considered part of the question which we discuss here. 

\begin{figure}[h]
    \centering
    \includegraphics[width=0.95\textwidth]{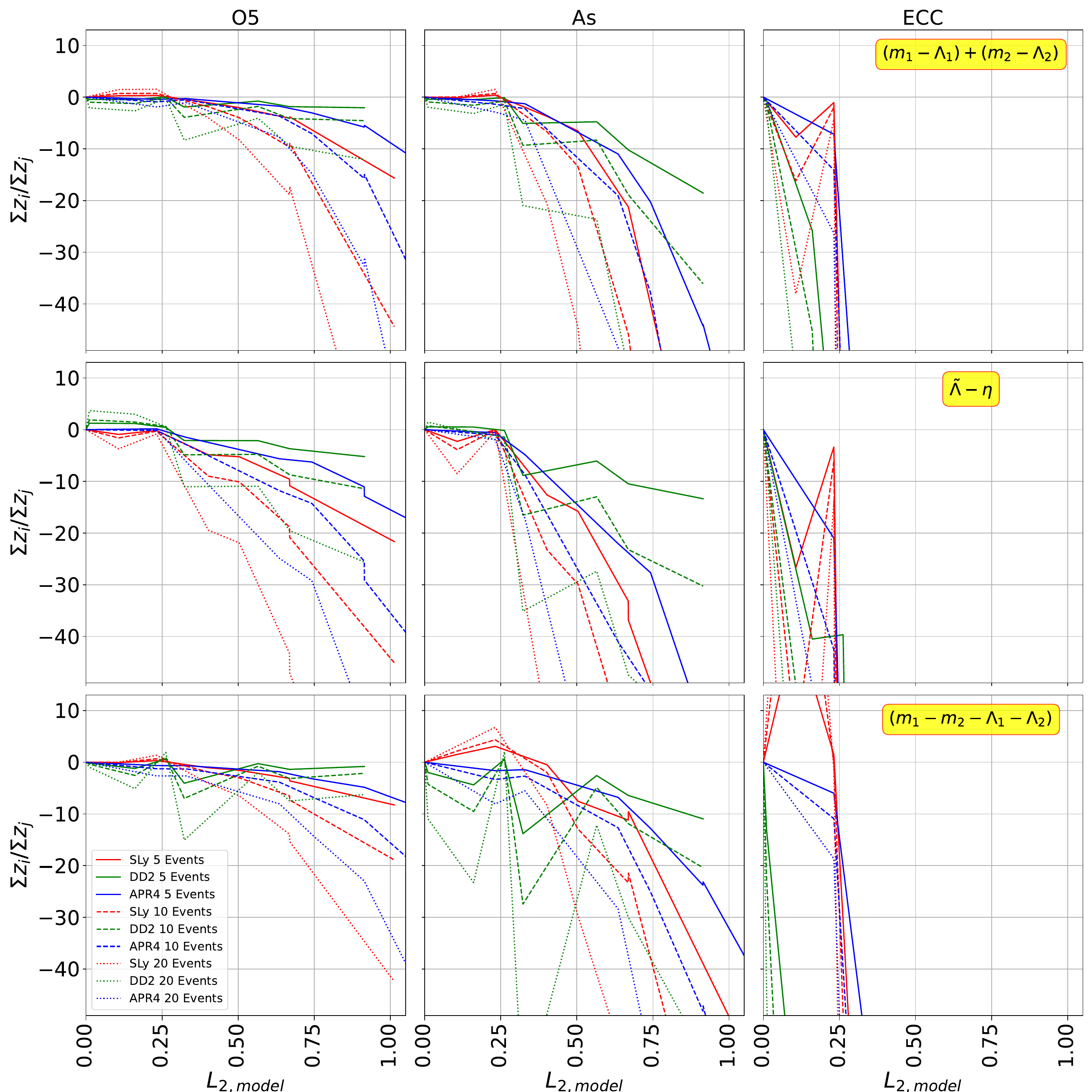}
    \caption{Cumulative Bayes Factor of EOS with respect to the injected EOS as a function of $L_{2,model}$ distance of all EOS models from the injeted EOS. The combined evidence is calculated using three methods, in two 2D spaces, $(m_1-\Lambda_1)$ and $(m_2-\Lambda_2)$, in 4D space, $(m_1-\Lambda_1-m_2-\Lambda_2)$ and in the $\eta-\tilde{\Lambda}$ space from top to bottom, respectively for each event. Please note that cutoff in the case of ECC is due to numerical underflow of evidence calculation due to very narrow posteriors distribution obtained typically for BNS PE runs for such detector configurations. This also results due to very narrow prior typically given in the relative binning framework of Bilby. Similar plots with other methods are shown in the appendix.}
    \label{fig:L2vsBF-all}
\end{figure}

We present the cumulative Bayes factor for the injected EOS with respect to the proximity of two EOS, that is, the distance $L_2$ as described in Section~\ref{subsubsec:EOS-choice}. We anticipate that when the chosen model EOS deviates significantly from the injected EOS, the accuracy of a particular method improves in identifying the correct EOS. Indeed, we find such a trend in call cases, which also addresses the issue of selecting the correct model in a function space. In Fig.~\ref{fig:L2vsBF-all}, we plot the Bayes factor of the models against the injected EOS quantified by the $L_2$ distance from the injected EOS, which gives a value 1 for the correct model and less than 1 for others. We find that as the EOSs move away from the injected model, the Bayes factor drops, with a decline rate that is proportional to the detector sensitivity and to the $L_2$-distance from the injected EOS. We observe small fluctuations in the Bayes factor due to the fact that the $L_2$-distance between two EOSs does not completely capture the difference between two functions representing the model. One may assume $BF>10$ as a criterion to prefer one model over another and $L_2>0.5$ to find the minimum number of events required to distinguish the models leading to up to $\sim 20$ events for O5, $\sim 10$ events for \As and $\sim5$ events for ECC.

For the case of O5 detectors, incorrect models can be ruled out only after observing 20 BNS mergers, whereas \As detectors would require only about 10 BNS events. For the ECC, the constraints on the EOS models are much better. Incorrect models can be distinguished with as few as 5 events. Due to sharply peaked posteriors, the evidence for models away from the injected EOS becomes numerically zero. Hence, optimistically, we can rule out all EOS models that are $L_2>0.25$ away from the "true" EOS model. This provides a method for constraining any falsifiable model of cold \ac{NS} matter.
 
\onecolumngrid

\twocolumngrid

We argue that such questions can be addressed on the basis of a population-informed model selection, which we defer to further studies. We elucidate this point with an example: DD2 and BHB share their $M$-$\Lambda$ curve for masses up to 1.6 M$\odot$ and differ for $M>1.6 M_\odot$. This means that we critically need larger NS masses in the BNS population to distinguish between them. Physically, it means that in the case of DD2 and BHB, the high-density EOS constraints are different, and constraints using GW observations require a special set of BNS events. A similar issue arises if two EOS intersect for some value of masses; in this case, the NS masses around the intersection point are not useful in distinguishing those two EOSs. Such pair is SLy and APR4 (see Fig.~\ref{fig:mr-curve}).

Moreover, we find that ignoring the pairwise correlation between
$p(m_1,\Lambda_1)$ and $p(m_2,\Lambda_2)$ significantly improves the selection
of the correct model. This is because such correlations do not contain
information about the EOS and ignoring this \ie avoid using the joint
posterior, $p(m_1,\Lambda_1,m_2,\Lambda_2)$ gives us the evidence in a smaller
parameter space where Bayesian evidence calculations do not suffer from Occam's
penalty. This approach offers a new lens through which we can view and understand the interconnections between different methods proposed in the literature on model selection using the parameter $\Lambda$. We emphasize that some correlations between BNS parameters are of astrophysical origin and are not determined by the EOS of NS matter, such as between mass and spin and between component masses. In addition, we point out that our current work is done for low-spin systems and hence mass-spin correlation measurement will be limited. We do not ignore the correlation between mass and tidal deformability, which is precisely what we aim to use and measure in our work. We aim to address correlations of astrophysical origin in our future research endeavors.

\subsection{Evidence for LIGO events: GW170817 and GW190425}

\begin{figure}[htb]
    \centering
    \includegraphics[width=0.5\textwidth]{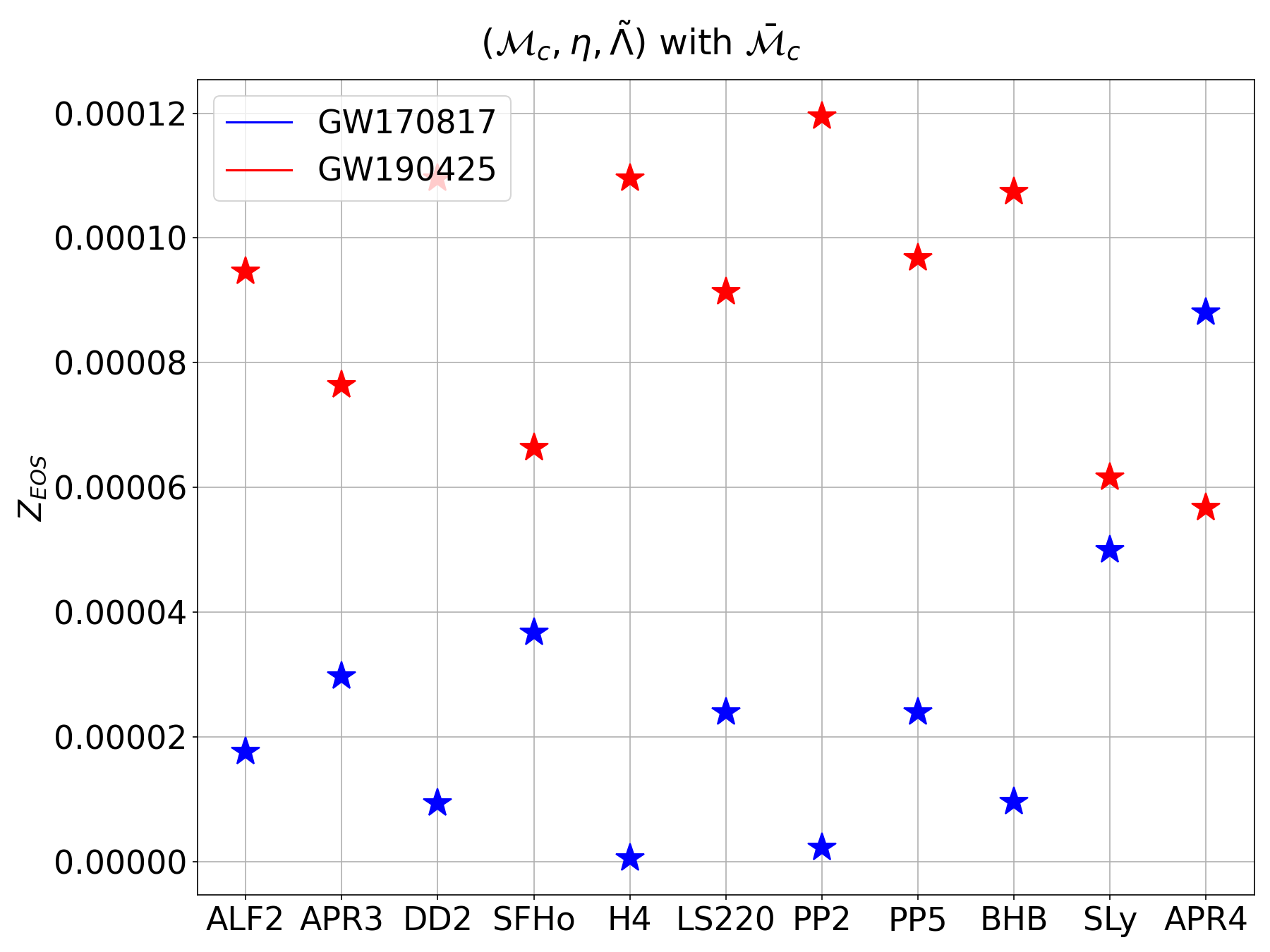}
    \caption{Evidence for GW170817 and GW190425 calculated for various EOSs in 2D space $\eta-\tilde{\Lambda}$}
    \label{fig:lvk-etalamt}
\end{figure}

\begin{figure}[htb]
    \centering
    \includegraphics[width=0.48\textwidth]{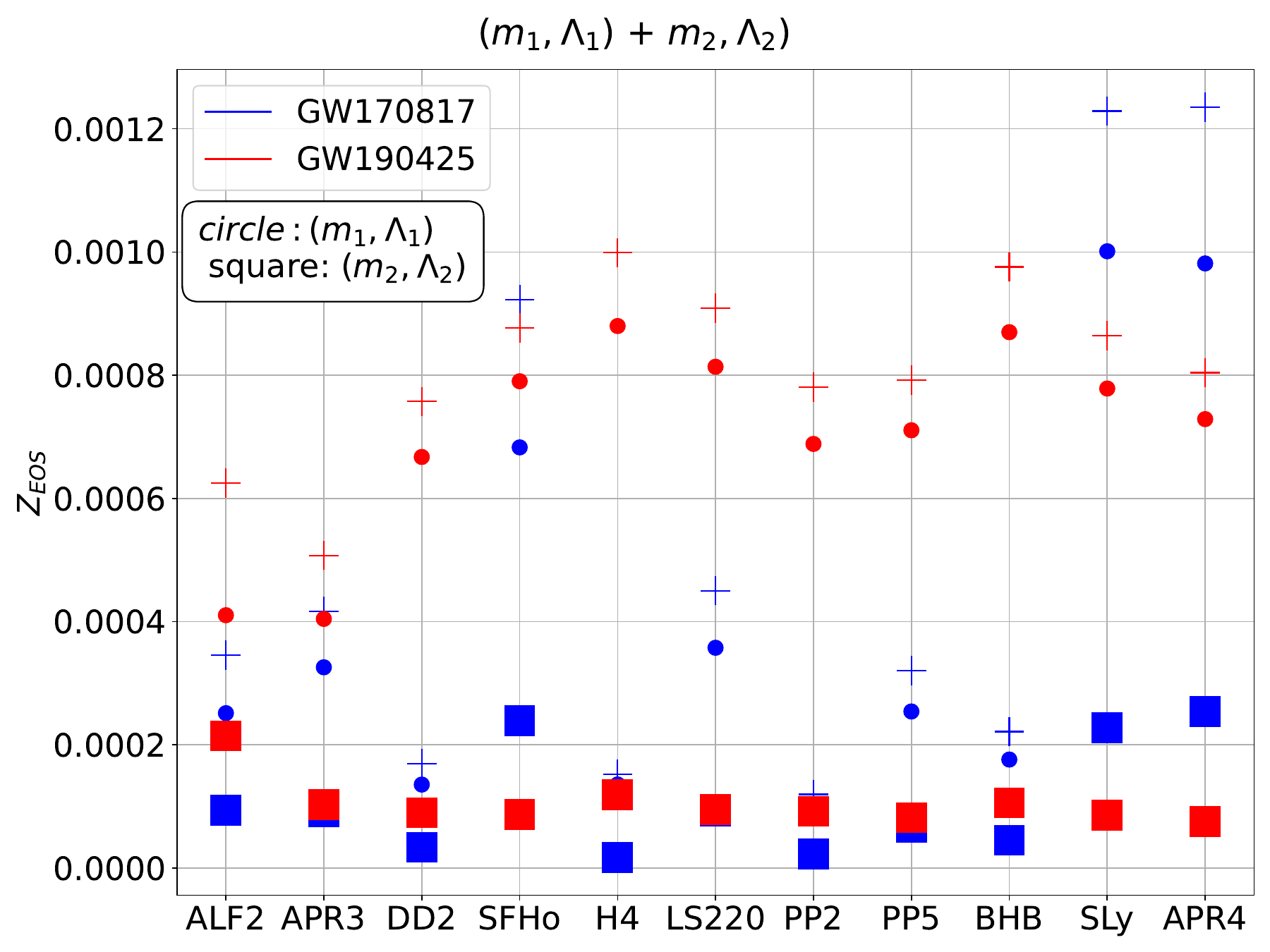}
    \caption{Evidence for GW170817 and GW190425 in the component $m-\Lambda$ subspace. The circle and square are evidence for two component masses and the $+$ represents the total evidence from the two events shown in color, red and blue.}
    \label{fig:lvk-mlam}
\end{figure}

We evaluate the evidence for two specific cases reported by the LIGO-Virgo Collaboration. These cases, identified as the BNS systems, were observed during O2 (GW170817) \citep{LIGOScientific:2018cki} and O3 (GW190425) \citep{Abbott2020-ge} observing runs. Utilizing the publicly available posterior samples, we computed the evidence using methods detailed in Sec.~\ref{sec:method}. The evidence in the $\tilde{\Lambda}$-$\eta$ parameter space is depicted in Fig.~\ref{fig:lvk-etalamt}, while Fig.~\ref{fig:lvk-mlam} illustrates the evidence in the $m$-$\Lambda$ parameter space.Our analysis indicates that in the $\tilde{\Lambda}$-$\eta$ parameter space, the data from GW190425 show a preference for the model PP2. On the other hand, when we consider the data from GW170817, the model APR4 is the one that stands out as the preferred model. Furthermore, when we change our perspective to the $m$-$\Lambda$ parameter space and closely scrutinize the evidence, it becomes apparent that APR4 is the preferred model based on data from GW170817. In contrast, the model H4 is preferred according to the posterior data obtained from GW190425.

Although our approach is robust enough to handle a collection of events, our injection study indicates that, given the current sensitivity of detectors, around 20 events are necessary to differentiate between different EoS models. Considering our current understanding of the BNS merger rate, we are fairly convinced that detecting 20 BNS events with an O5 run of the LVK is unlikely. Nevertheless, certain sets of EoSs that exhibit significant differences (in terms of $L_2$ norms) can be excluded. However, this still requires approximately up to 10 events.

\section{Conclusions and Summary}
\label{sec:conclusions}

In this study, we present a general Bayesian framework discriminating between EOSs of cold \ac{NS} matter that are given in a tabular form\footnote{The \ac{BEOMS} pipeline will be made public as a Python's \texttt{pip} package in the future.}. In this format, our formalism applies to any relation describing a model where the independent variable is the NS mass and the dependent variable may consist of one or more of a list of parameters including tidal deformability, radius, spin-induced quadrupole constant \citep{Laarakkers:1997hb}, or any other property that can impact the waveform and can be constrained by GW observations.

A single \ac{EOS}-agnostic \ac{PE} run is performed for each \ac{BNS} event to compute the evidence and subsequently rank the EOS models based on the accumulated evidence from a randomly selected ensemble of events. We calculate the evidence using the 4D posterior PDF where both masses ($m_1,m_2$) and tidal deformabilities ($\Lambda_1, \Lambda_2$) are sampled independently. Here, the functional dependences (i.e. EOS) of both masses on tidal deformabilities are assumed to be the same. Subsequently, the PDF is transformed into the coordinate system defined by $(\mathcal{M},\eta,\tilde{\Lambda})$, and the evidence is recalculated under the assumption of identical EOS for both components. In both parameter spaces, the presence of mass-ratio data within the posterior PDF introduces degeneracies, thereby diminishing the efficacy of the model selection process. A novel approach is proposed: The exclusion of data on the mass ratio enables the computation of evidence within the $m$-$\Lambda$ space by employing two marginalized PDFs, $p(m_1,\Lambda_1)$ and $p(m_2,\Lambda_2)$. This methodology appears to be more suitable for the EOS model selection, particularly when analyzing inspiral GW signals from BNS mergers. Additionally, the Bayes factors for various EOS models are computed and compared against the injection EOS using the aforementioned strategies.

For detectors with O5 sensitivity, evidence calculations across parameter space reveal no meaningful differences between the models in question. However, for the \As and ECC networks, performing evidence calculations in the subspace $m$-$\Lambda$ is more effective in identifying the correct model than in other spaces. This is because excluding the mass ratio from evidence calculation does not eliminate any details concerning the \ac{EOS} of nuclear matter. Additionally, this choice reduces the dimensional complexity of the evidence calculation.

We validate and demonstrate the robustness of this hypothesis through extensive injection studies. This is particularly relevant for next-generation detectors with enhanced sensitivity, such as ECC, where we can eliminate ambiguities related to the mass ratio and the \ac{EOS}. This approach is analogous to employing universal relations that also involve the mass ratio. These universal relations proved to be effective, or perhaps equivalently ineffective, for detectors with lower sensitivity because the systematic errors resulting from the assumed universality are significantly smaller than the statistical errors.

Our main conclusions are as follows. 
\begin{itemize}
    \item We rigorously evaluate the methods of evidence calculation for an EOS model across various spaces spanned by the parameter space of BNS mergers and their respective subspaces to demonstrate the relative efficiency in identifying the accurate model.
    \item The evidence of an EOS model increases with respect to the distance between two EOSs as described by the $L_2$ distance between the models.
    \item The calculations of evidence in subspaces demonstrate an overall higher model-selection efficiency. We recommend this approach as one of the primary methods for subsequent analyzes. However, we note that such an efficiency also depends on the population of BNS mergers analyzed, which we leave for future studies.
\end{itemize}

We highlight that selecting a fiduacial NS mass distribution for injection study based on an EOS could skew the results due to problems with overlapping and intersecting EOS models as outlined in Sec.~\ref{subsubsec:EOS-choice}. We show that for EOSs that overlap and intersect, population-informed model selection is necessary to identify and differentiate the variations in EOSs present in contemporary theoretical frameworks. We will address these issues in subsequent studies.

\begin{acknowledgments}
We extend our appreciation to Geraint Pratten, Bikram Pradhan, and Justin Janquart for their meticulous review of the \ac{BEOMS} code as well as their invaluable feedback. The authors are grateful to Parameswaran Ajith for starting the conversation and supporting us in addressing this issue. We also thank Rossella Gamba, Koustav Chandra, and Nathan Johnson McDaniel for their thoughtful feedback and commentary. Our work utilizes the LVK computing resources located at Penn State. This material is based on work supported by the National Science Foundation's LIGO Laboratory, which is a major facility fully funded by the National Science Foundation. We also acknowledge the National Science Foundation support via OAC-2346596, OAC-2201445, OAC- 2103662, OAC-2018299, PHY-2110594.
\end{acknowledgments}

\bibliographystyle{unsrtnat}
\bibliography{eosBayesianModelSelection, refsxgraderr}

\acrodef{BEOMS}{Bayesian Evidence calculation fOr Model Selection}

\acrodef{ADM}{Arnowitt-Deser-Misner}
\acrodef{AMR}{adaptive mesh-refinement}
\acrodef{BH}{black hole}
\acrodef{BBH}{binary black-hole}
\acrodef{BHNS}{black-hole neutron-star}
\acrodef{BNS}{binary neutron star}
\acrodef{BWD}{binary white dwarf}
\acrodef{CBC}{compact binary coalescence}
\acrodef{CCSN}{core-collapse supernova}
\acrodefplural{CCSN}[CCSNe]{core-collapse supernovae}
\acrodef{CMA}{consistent multi-fluid advection}
\acrodef{DG}{discontinuous Galerkin}
\acrodef{HMNS}{hypermassive neutron star}
\acrodef{ECSS}{Extended Collaborative Support Service}
\acrodef{EM}{electromagnetic}
\acrodef{ET}{Einstein Telescope}
\acrodef{EOB}{effective-one-body}
\acrodef{EOS}{equation of state}
\acrodef{FF}{fitting factor}
\acrodef{GR}{general-relativistic}
\acrodef{GRB}{$\gamma$-ray burst}
\acrodef{GRLES}{general-relativistic large-eddy simulation}
\acrodef{GRHD}{general-relativistic hydrodynamics}
\acrodef{GRMHD}{general-relativistic magnetohydrodynamics}
\acrodef{GR-nuR-MHD}{general-relativistic neutrino and radiation magnetohydrodynamics}
\acrodef{GW}{gravitational wave}
\acrodef{Ho}{Hubble-Lemaitre constant}
\acrodef{ILES}{implicit large-eddy simulations}
\acrodef{KDE}{kernel density estimation}
\acrodef{LIA}{linear interaction analysis}
\acrodef{LES}{large-eddy simulation}
\acrodef{LIGO}{Laser Interferometer Gravitational Wave Observatory}
\acrodef{MHD}{magnetohydrodynamics}
\acrodef{MRI}{magnetorotational instability}
\acrodef{NR}{numerical relativity}
\acrodef{NS}{neutron star}
\acrodef{PDF}{probability density function}
\acrodef{PE}{parameter estimation}
\acrodef{PN}{post-Newtonian}
\acrodef{PNS}{protoneutron star}
\acrodef{RMHD}{radiation hydrodynamics}
\acrodef{SASI}{standing accretion shock instability}
\acrodef{SGRB}{short $\gamma$-ray burst}
\acrodef{SKA}{Square Kilometer Array}
\acrodef{SN}{supernova}
\acrodefplural{SN}[SNe]{supernovae}
\acrodef{SNR}{signal-to-noise ratio}
\acrodef{TMT}{Thirty Meter Telescope}
\acrodef{TOV}{Tolman-Oppenheimer-Volkoff}


\appendix 

\onecolumngrid

\section{Evidence distribution for APR4 and SLy}

\begin{figure*}[ht]
    \centering
    \includegraphics[width=0.87\textwidth]{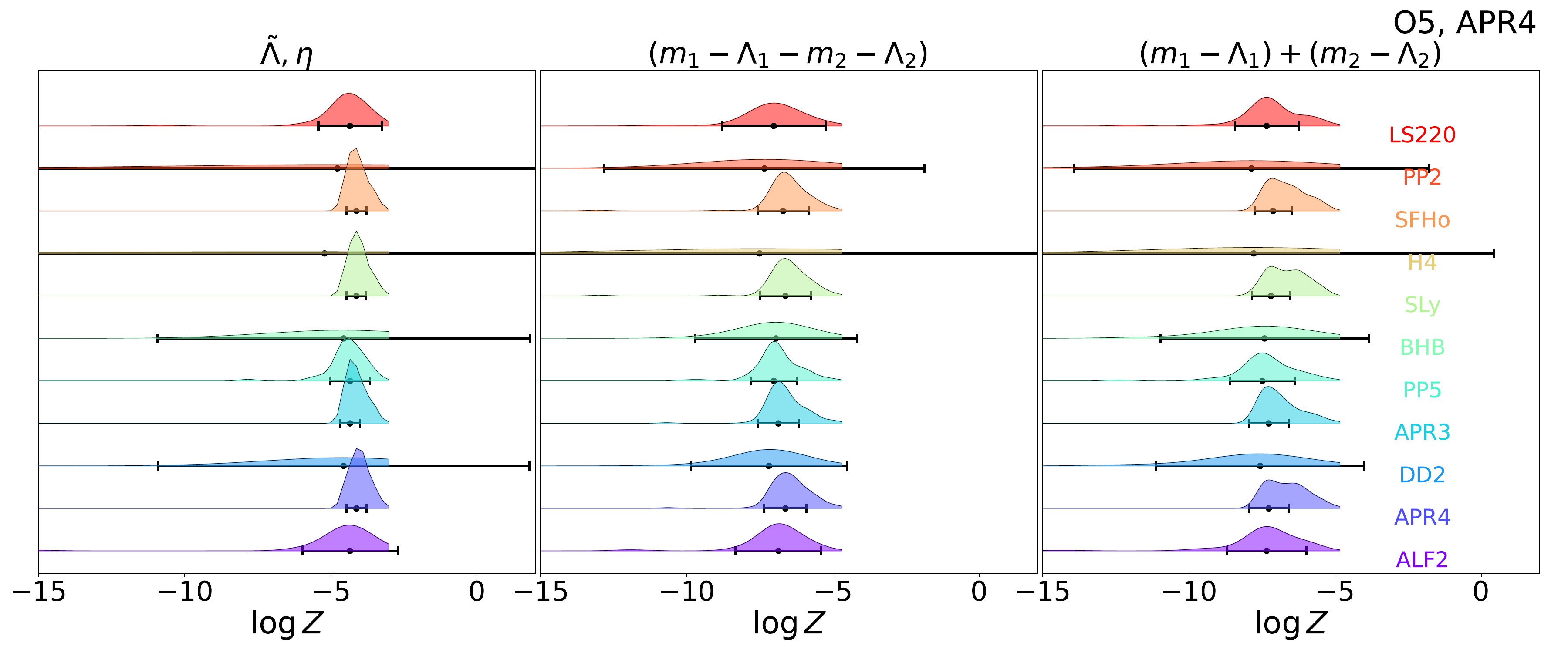}
    \includegraphics[width=0.87\textwidth]{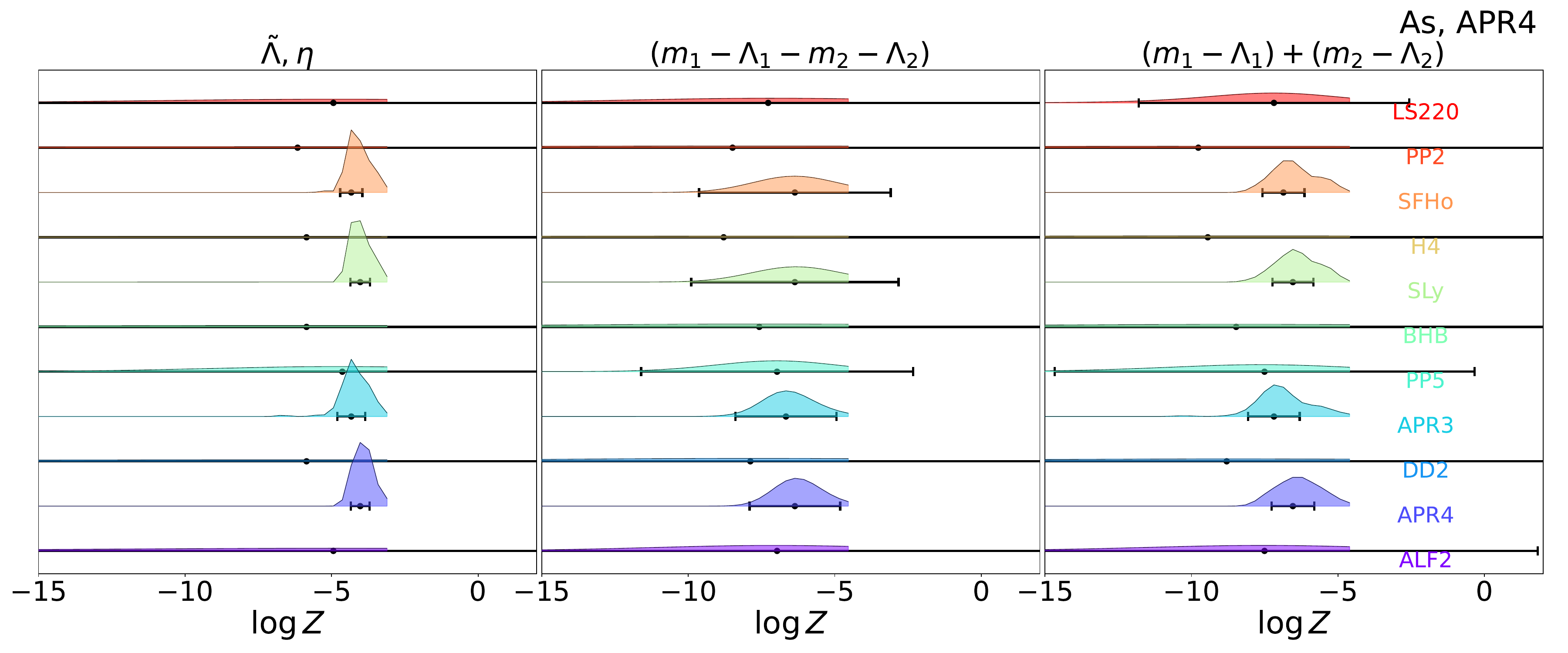}    
    \includegraphics[width=0.87\textwidth]{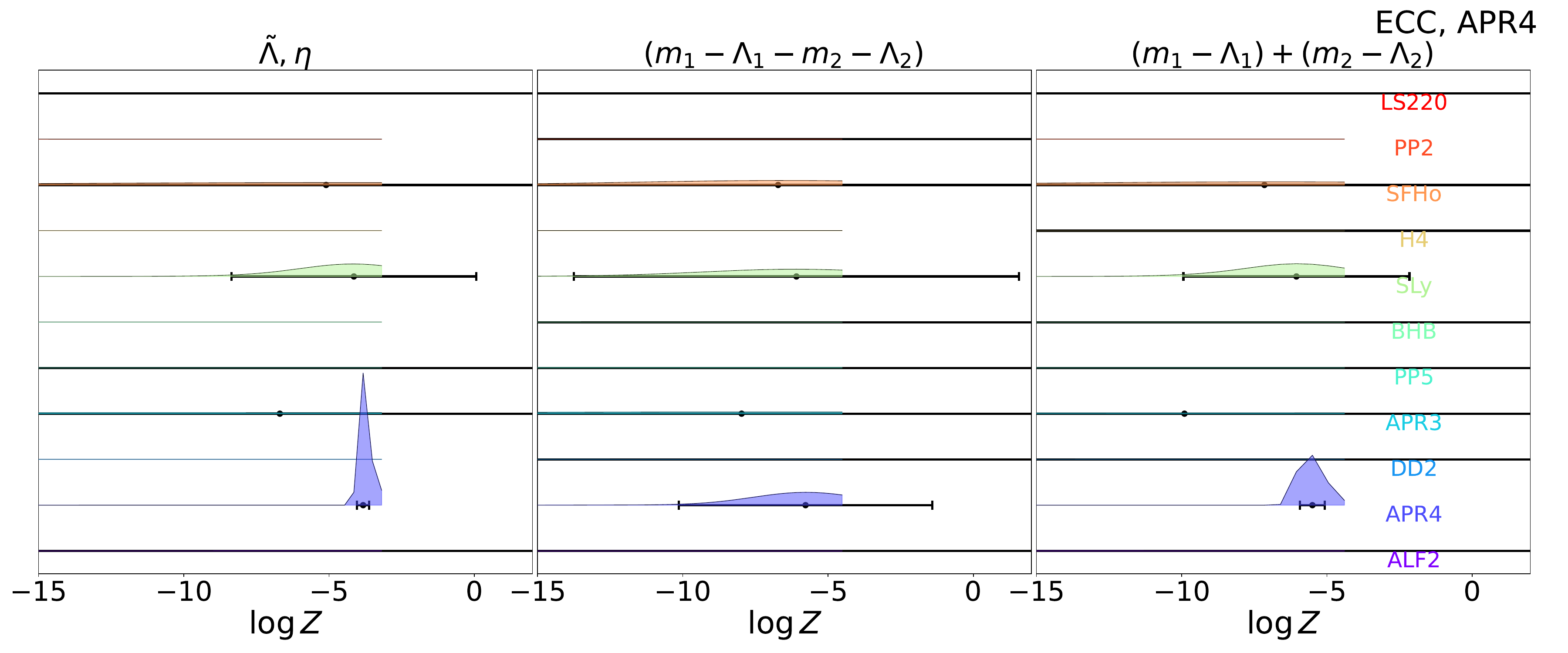}
    \caption{This is similar to Fig.~\ref{fig:distEvidence-DD2} but for injection of EOS, APR4. We find that APR4 and SLy being closest has similar evidences.}
    \label{fig:distEvidence-APR4}
\end{figure*}

\begin{figure*}[ht]
    \centering
    \includegraphics[width=0.87\textwidth]{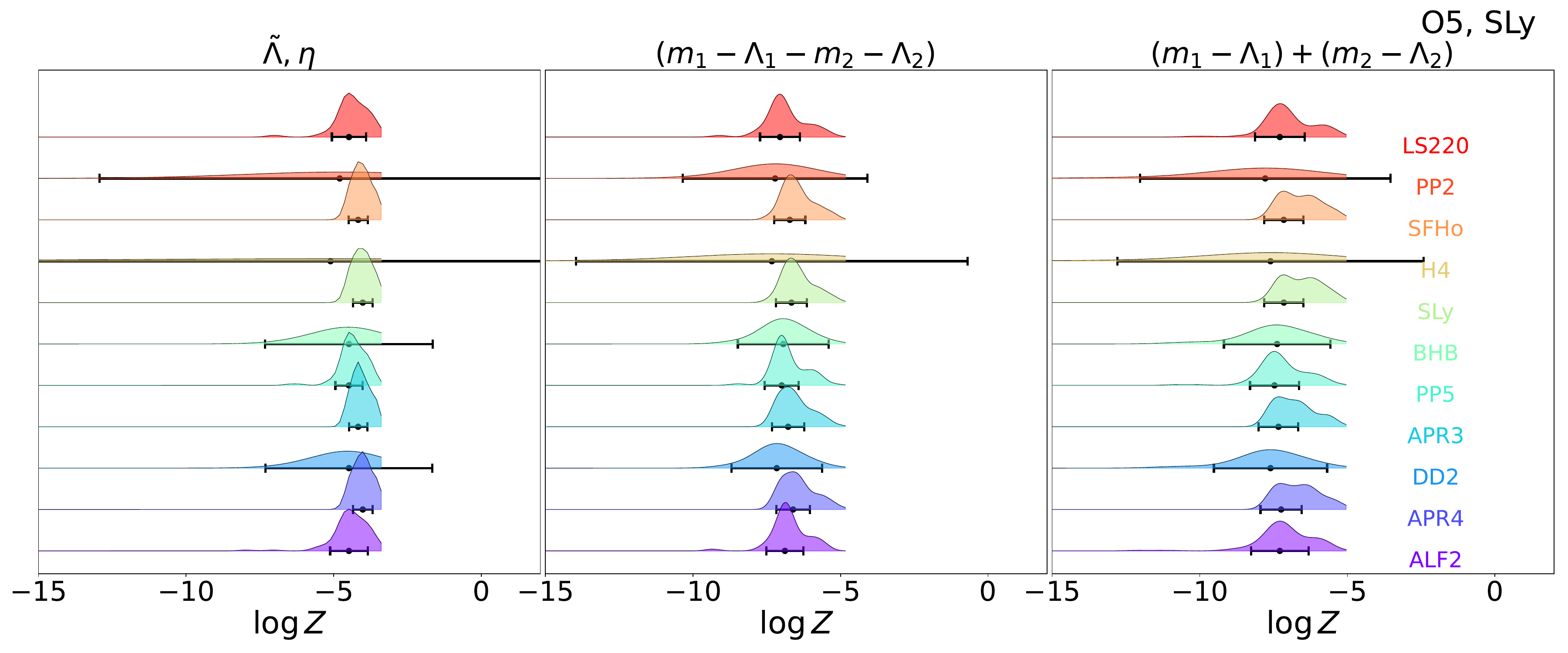}
    \includegraphics[width=0.87\textwidth]{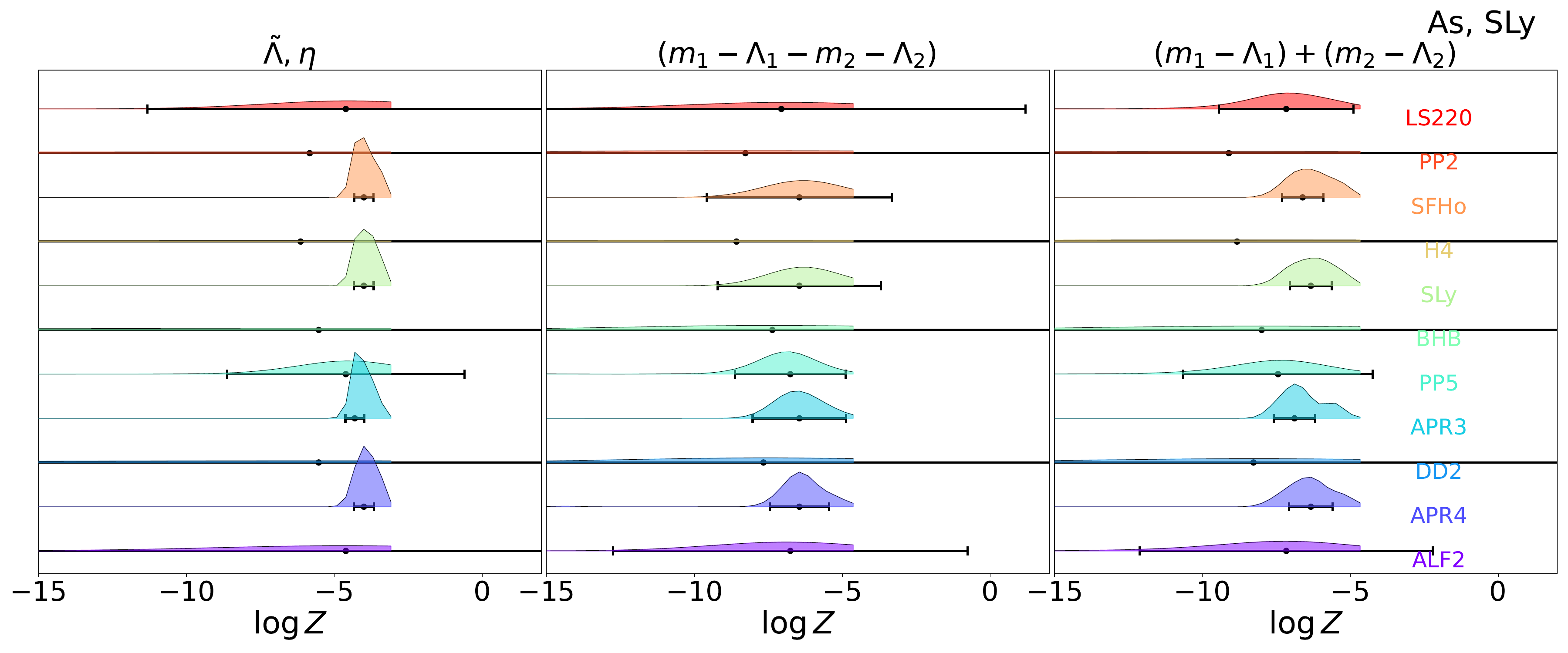}    
    \includegraphics[width=0.87\textwidth]{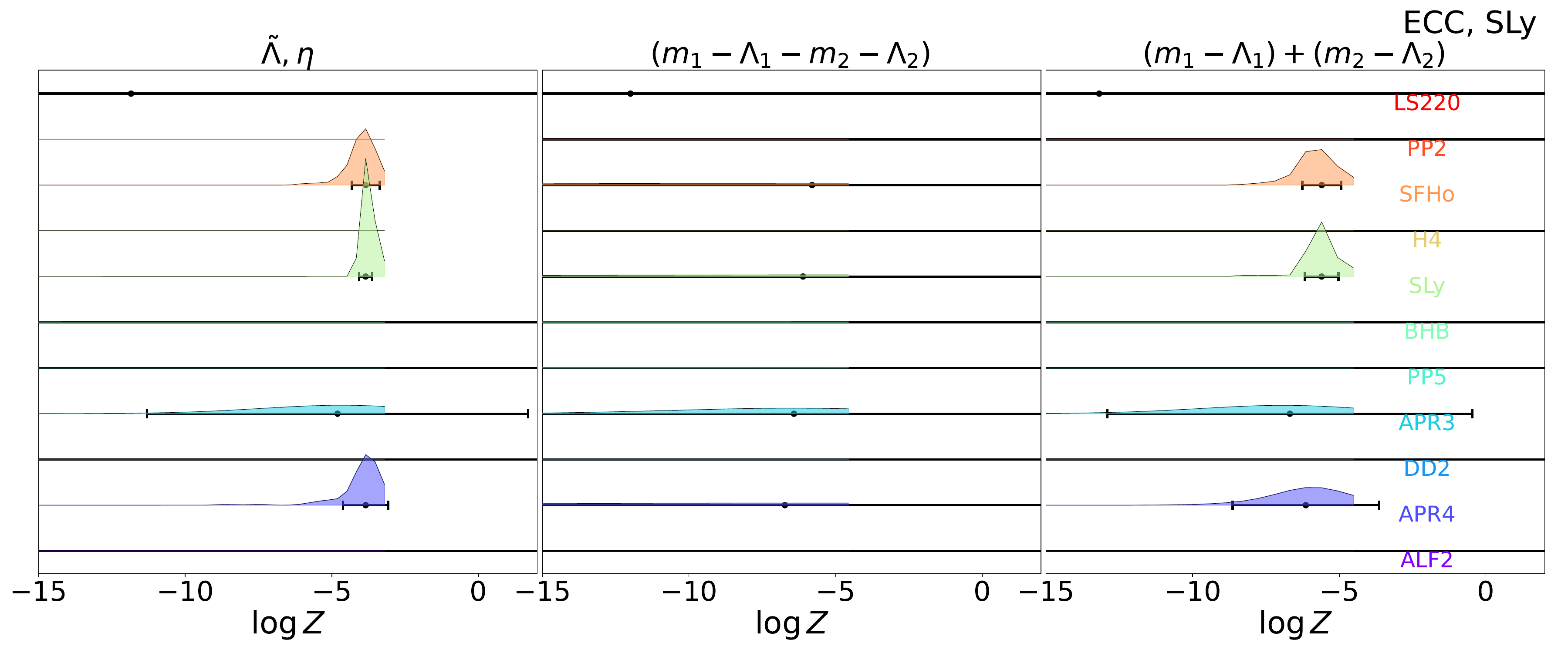}
    \caption{This is similar to Fig.~\ref{fig:distEvidence-DD2} but for injection of EOS, SLy. In this figure, two other EOS, SFHo and APR4, obtain closer evidence values but with broader evidence compared to SLy due to their nearness in $(m-\Lambda)$ space (see Fig.~\ref{fig:mr-curve}).}
    \label{fig:distEvidence-SLy}
\end{figure*}

This section extends the description of the evidence distribution for equation of state (EOS) models, specifically APR4 and SLy, as outlined in the main text. It is noted that there are degeneracies with other EOS models, which become apparent with the most sensitive detectors. The evidence for APR4 and SLy models is found to be similar. The O5 detector is capable of excluding some more remote models such as H4 and PP2. Moreover, with configuration \As, it is possible to exclude DD2, LS220, BHB, PP5, and ALF2. When APR4 is used as the injection EOS (refer to Fig.~\ref{fig:distEvidence-APR4}), the ECC is able to distinctly identify it, although there is a moderate level of degeneracy with the SLy model. Conversely, when SLy is employed as the injection EOS (refer to Fig.~\ref{fig:distEvidence-SLy}), the resulting degeneracies are somewhat more pronounced, with SFHo emerging as a close contender. It is observed that the 4D evidence calculation does not accurately identify the correct model and instead incorrectly favors another model, as illustrated in Fig.~\ref{fig:L2vsBF-all}.

\section{Definition of $L_2$}\label{apdx:L2}

The $L_2$-distance between the mass-radius or mass-$\Lambda$ curves can be used as a measure of the distance between a pair of EoS models. We propose that two models are more easily distinguishable greater is the $L_2$-distance between the corresponding curves defined as: 

\begin{equation}
    L_{2,\Lambda}(A,B) \equiv N_\Lambda \int_{m_l}^{m_u} [\Lambda_A(m)-\Lambda_B(m)]^2 dm 
\end{equation}
where  $N_\Lambda$ is normalization constants to render the distances dimensionless chosen to be:
\begin{eqnarray}
    N_\Lambda \equiv \left [ \int_{m_l}^{m_u} \Lambda_A^2 dm \int_{m_l}^{m_u} \Lambda_B^2 dm \right ]^{-1/2}
\end{eqnarray}
 $\Lambda_A(m)$ and $\Lambda_B(m)$ are the mass-$\Lambda$ curves corresponding to model $A$ and $B$ respectively, $m_l$ is the smallest NS mass in the observed population and $m_u$ is the smaller of the maximum mass allowed by models $A$ and $B.$

\section{KDE Reconstruction check}
	
\begin{figure}[h!]
	\centering
	\includegraphics[width=0.8\textwidth]{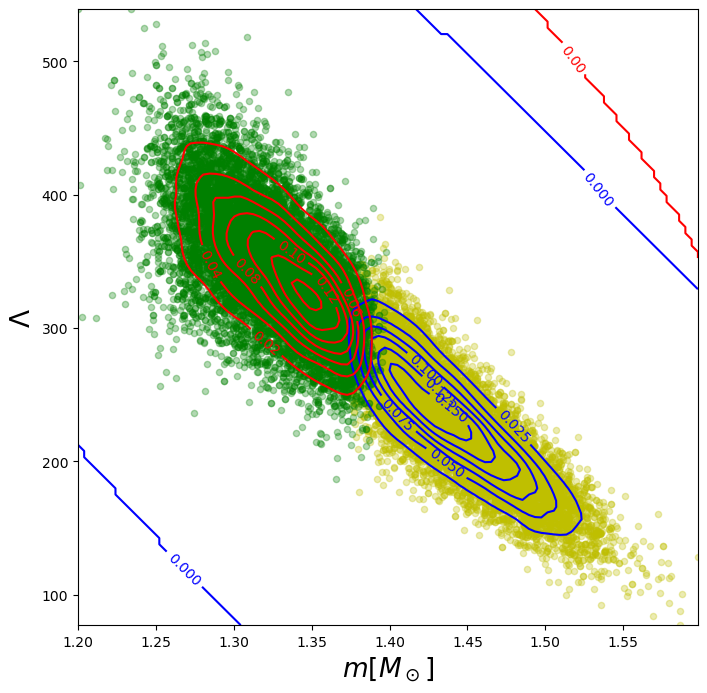}
	\caption{Comparison of the KDE-reconstructed posterior (contours) and the scatter plot of samples (circles) for one representative case, EOS being APR4 and detector configuration being ECC. The KDE adequately captures the structure visible in the raw samples.}
	\label{fig:kde_scatter}
\end{figure}

\end{document}